\documentclass[]{article}

\usepackage{inputenc}
\usepackage{graphicx}
\usepackage{hyperref}
\usepackage{booktabs}
\usepackage{times}
\usepackage{graphicx}
\usepackage{dcolumn} %
\usepackage{bm}      %

\usepackage{caption}
\usepackage{subcaption}

\usepackage{cite}
\usepackage[usenames]{color}
\usepackage{rotating}
\usepackage{fontenc}
\usepackage{amsthm}
\usepackage{hyperref}
\usepackage{cancel}
\usepackage{rotating}
\usepackage[dvipsnames]{xcolor}
\usepackage{authblk}
\usepackage{epstopdf}
\usepackage[]{algorithm2e}
\usepackage{enumerate}
\usepackage{enumitem}

\usepackage{comment}
  
\usepackage{array}
\usepackage{amssymb}
\usepackage{amsfonts}
\usepackage{amsmath}
\usepackage{mathrsfs}

\usepackage{graphicx}        
\DeclareGraphicsExtensions{.pdf,.jpeg,.png,.jpg,.eps} 
\usepackage{hyperref}
\usepackage{float}
\usepackage[euler]{textgreek}

\usepackage{color}  
\definecolor{purple}{rgb}{1.0, 0.0, 1.0}

\title{Synaptic Time-Dependent Plasticity Leads to Efficient Coding of Predictions}
\author[1,2]{Pau Vilimelis Aceituno}
\author[1]{Masud Ehsani}
\author[1,3]{J\"{u}rgen Jost}
\affil[1]{Max Planck Institute for Mathematics in the Sciences, Inselstra{\textbeta}e 22, 04103 Leipzig, Germany}
\affil[2]{Max Planck School of Cognition, Stephanstra{\textbeta}e 1a, 04103 Leipzig}
\affil[3]{Santa Fe Institute, 1399 Hyde Park Road, 87501 Santa Fe, New Mexico, USA}

\date{}
\begin{document}
	
	\maketitle
	
	\begin{abstract}
		Latency reduction of postsynaptic spikes is a well-known effect of Synaptic Time-Dependent Plasticity. We expand this notion for long postsynaptic spike trains, showing that, for a fixed input spike train, STDP reduces the number of postsynaptic spikes and concentrates the remaining ones. Then we study the consequences of this phenomena in terms of coding, finding that this mechanism improves the neural code by increasing the signal-to-noise ratio and lowering the metabolic costs of frequent stimuli. Finally, we illustrate that the reduction of postsynaptic latencies can lead to the emergence of predictions. 
	\end{abstract}

	\section{Introduction}
	Living organisms need to make accurate predictions in order to survive \cite{bubic2010prediction,hohwy2013predictive}, posing the question of how do brains learn to make those predictions. Early general models based on classical conditioning \cite{rescorla1972theory,miller1995assessment}, as well as mechanistic models explaining the neural substrate for those predictions \cite{schultz1997neural,heeger2017theory} assume that prediction performance is feed back into the predicting neural population, similarly to supervised or reinforcement learning paradigms that are common in machine learning. However, recent studies have found that sensory neurons without feedback from higher brain areas encode predictive information \cite{palmer2015predictive}, a finding that has been supported by simulation studies \cite{sederberg2018learning}. This implies that a bottom-up process without explicit feedback --similar to unsupervised learning-- should also generate predictions. 
	
	In this paper we present such a mechanism by focusing on postsynaptic latency reduction. This is a well-known effect of Synaptic Time-Dependent Plasticity (STDP) first mentioned by Song et. al \cite{song2000competitive} for a single post-synaptic neuron driven by an specific excitatory input pattern. This effect was explored in detail in a simulation study by Guyonneau et. al \cite{guyonneau2005neurons} who showed that this effect is robust to noise in the form of jitter and Poissonian spontaneous activity. They further analyze STDP on a single neuron receiving fixed (among trials) Poissonian spike trains from each pre-synaptic neuron and showed that by STDP weights of the earliest afferents will be increased, regardless of the firing rate and level of synchrony of the corresponding neurons. Masquelier et. al \cite{masquelier2008spike} showed how a single post-synaptic neuron by  STDP would learn a single frequent excitatory pattern of spikes even when there is a strong background noise presented and associate this with how postsynaptic firing latency in response to pattern presentation would decrease. In this article we revisit this phenomenon at the micro level with plastic inhibitory neurons added to the previous setups and analyze the effect of latency reduction at network level.
	
	The gist of our argument is that latency reduction implies that neurons fire as early as possible for a given input spike train that is repeated very often; as neurons do not differentiate between a specific stimulus and an early clue of such a stimulus -- both being part of seemingly the same input spike train-- the neurons can, by STDP, fire earlier than the stimulus itself. Furthermore, we expand on the previous studies focused on excitatory neurons to include inhibition and illustrate the parameter regime in which inhibitory plasticity is compatible with latency reduction. However, the latency reduction mechanism has other uses in terms of neural code. First, as neurons fire as early as possible when a stimulus is presented, their spikes will concentrate in a small time window, and thus they are easier to decode. Second, we show that the latency reduction can also lead to a reduction of the number of spikes, which translates as a reduction of metabolic costs for encoding frequent stimuli. 
	
	We develop our argument by studying simple models of neurons subject to fixed input spike trains. We use a combination of simulations and mathematical analysis to derive our results, starting from the evolution of a single postsynaptic spike at very short timescales we expand to larger scales that conclude in the emergence of predictions and efficient code at the level of populations of neurons in large timescales.
	
	The rest of this paper is organized as follows. First, we present the models of neurons and STDP in Sec.~\ref{sec:Models}. Second, study the effects of STDP in a single postsynaptic spike in very small timescales $\sim 10 ms$, focusing on latency reduction and the reduction of the number of postsynaptic spikes in Sec.~\ref{sec:shortTimescales}. Third, we expand those results to long postsynaptic spike trains in Sec.~\ref{sec:postSpikeTrain}, finding that STDP forces postsynaptic neurons to fire only once at the onset of the presynaptic spike train. Third, we provide an  interpretation of this concentration in terms of neural code performance, showing that it leads to lower metabolic costs and lower the decoding errors in Sec.~\ref{sec:EncodingEffects}. We finalize by illustrating that the same mechanism of latency reduction leads to encoding predictions in Sec.~\ref{sec:Predictions}.
	
	\section{Models}\label{sec:Models}
	
	\subsection{Leaky integrate-and-fire neuron}
	
	Neurons are cells that constitute the basic computational units in the nervous system. Their main feature is the capacity to receive information through electrical impulses, combine this information and send impulses to other neurons. In this paper we model them as Integrate-and-Fire neurons with a refractory period \cite{IFModel}. In this model, the state of a neuron at a given time is described by its membrane potential $v(t)$, which evolves according to the equation
	\begin{equation}\label{eq:LIF}
	\tau_m \dfrac{d v(t)}{dt} = -(v(t) - v_0) + i(t),
	\end{equation}
	where $\tau_m = 10 ms$, $v_0 = -70 mV$. $i(t)$ is the input to the neuron at time $t$. When the membrane potential reaches a certain threshold $v_{th} = -50 mV$, the neuron "fires" or "spikes", meaning that it emits a pulse of current. After firing, the membrane potential is reset to its resting state $v_0$ and kept frozen at this value for a fixed period of time called the refractory period $t_{ref} = 4 ms$.
	
	The firing of a neuron generates pulses of current that arrive at other neurons, which in turn update their membrane potentials. If neuron $a$ receives the spikes of neuron $b$ we will say that there is a synapse going from the second to the first. The receiving neuron is called postsynaptic and the sending neuron is the presynaptic one. This synapse is characterized by a weight $w_{ab}$ and a delay $d_{ab}$ which correspond, respectively, to the gain and the latency that the pulse of neuron $a$ goes through before arriving at $b$. 
	
	\subsection{Input spike trains}
	
	Neurons communicate mainly through action potentials or spikes, which are typically modeled as Diracs, hence the input to a neuron can be described as
	\begin{equation}\label{eq:LIT_input}
	i(t) = \sum_n w_n\delta(t-t_n),
	\end{equation}
	where $w_n$ is the weight of the spike, which corresponds to the strength ot the synapse from which the spike comes, and $t_n$ is the arrival time of the spike. The weights of the synapses can be positive, if the presynaptic neuron is excitatory, or negative, if it is inhibitory. Through this paper we will assume that every neuron gets an input that will be repeated, meaning that a neuron will always get spikes from different synapses, and although the weights of the synapses might change, the times $t_n$ of the spikes will remain the same in every input repetition.

	\subsection{Synaptic Time-Dependent Plasticity}
	
	Networks of neurons learn by modifying the strength of the connections between them --previously denoted by $w$ There is a rich literature on what rules those weights follow in biological neurons and their respective implications \cite{dayan2001theoretical}. For the purposes of this paper, the biological neurons that we will analyze and simulate will adapt their connections according to the Synaptic Time-Dependent Plasticity (STDP) paradigm\cite{STDPScholarpedia,gerstner1996neuronal}.
	
	In STDP the weight of a connection is modified depending on the time interval between pairs of pre- and post-synaptic spikes. For every pair the weight of the synapse is modified according to the equations
	\begin{equation}\label{eq:STDP_1}
	\Delta w (\Delta t) = \begin{cases}
	A_{+}(w) e^{-\frac{|\Delta t|}{\tau_{s}}}
	\ \ if \Delta t \geq 0
	\\
	-A_{-}(w) e^{-\frac{|\Delta t|}{\tau_{s}}}
	\ \ if \Delta t < 0
	\end{cases}
	\end{equation}
	where $\Delta t = t_{post} - t_{pre}$ is the time difference between the postsynaptic spike and the presynaptic one, $\tau_{s} = 20 ms$. Based on previous works \cite{kistler2000modeling,van2000stable}, we define $A_+$ and $A_-$ as
	\begin{equation}\label{eq:STDP_2}
	\begin{aligned} 
	A_+(w) &= \eta_+ (w_{\max}-w),
	\\
	A_-(w) &=  \eta_- (w-w_{\min})
	\end{aligned}
	\end{equation}
	where $\eta_- = 0.015,\ \eta_+ = 0.01$, $w_{\max}^e = 10 mV$ and $w_{\min} = 0$. Inhibitory synapses follow the same rules as their excitatory counterparts but with parameters $\eta_- = 0.045,\ \eta_+ = 0.03$ and $w_{\max}^i = 20 mV$. Each synapse comes from a presynaptic excitatory neuron with probability $0.8$ or inhibitory with probability $0.2$. 
	
	\subsection{Model limitations and required features}
	
	We must note that the models used here are heavy simplifications of real neurons. LIF neurons do not exhibit the rich range of dynamics that real neurons can have \cite{izhikevich2004model}, ion channel kinetics are more complicated than simple Diracs \cite{chapeau1995synapse} and the STDP model used here cannot account for the evolution of synaptic weights when the frequency of postsynaptic or presynaptic spikes is high \cite{pfister2006triplets}. However, those models contain the main biologically realistic features that we need for the rest of this study. First, the time constants of the neuron membrane potentials\cite{gerstner2014neuronal} and the STDP interactions \cite{bi1998synaptic} are at least an order of magnitude smaller than the duration of the input spike trains associated to biologically realistic stimuli-evoked spatiotemporal patterns\cite{rolston2007precisely,prut1998spatiotemporal}. Second, the sparse firing of the typical neuron \cite{roxin2011distribution}. Third, the potentiation of presynaptic spikes preceding postsynaptic ones, which is consistent with most experiments \cite{pfister2006triplets}. Finally, the homeostatic consideration that neurons should not widely increase their firing rate, which is a natural requirement on metabolic grounds \cite{turrigiano2004homeostatic} can easily be incorporated by the depressive term $A_-$. Thus we will keep these well-known models \cite{gerstner2014neuronal} on the grounds that they are analytically tractable and qualitatively plausible.

	\section{Evolution of a single postsynaptic spike}\label{sec:shortTimescales}
	
	In this section we show that STDP can change individual postsynaptic spikes by reducing their latencies and their number. We will start by presenting simple scenarios with excitatory inputs in which both effects are easy to illustrate, then show how inhibitory synapses can be added to the model, and finally show that those effects can appear in random input spike trains by presenting simulations. It is worth noticing that the time windows in this section are on the order of $\tau_s$ and the number of repetitions of each input pattern will be small. 
	
	\subsection{Latency Reduction}
	
	If a fixed train of presynaptic spikes is repeated very often, then the spikes that arrive before the postsynaptic spike get reinforced. This implies that the postsynaptic spike might then be triggered earlier \cite{song2000competitive,gerstner1996neuronal}. When this happens, the refractory period of the postsynaptic neuron would prevent a second spike on the original spiking site. However, when the postsynaptic spike happens earlier and earlier, it might lead to a proliferation of spikes by having a new spike appear at the time of the original postsynaptic spike. Following previous literature \cite{song2000competitive,abbott2000synaptic,kempter2001intrinsic}, to prevent this effect, we assume that long term depression -- the weakening of synaptic weights -- is stronger than long term potentiation -- the strengthening of postsynaptic weights.  
	
\begin{figure}
		\begin{subfigure}{0.48\textwidth}
			\centering
			\includegraphics[width=1\linewidth,trim={0cm, 0cm, 0cm, 0cm},clip]{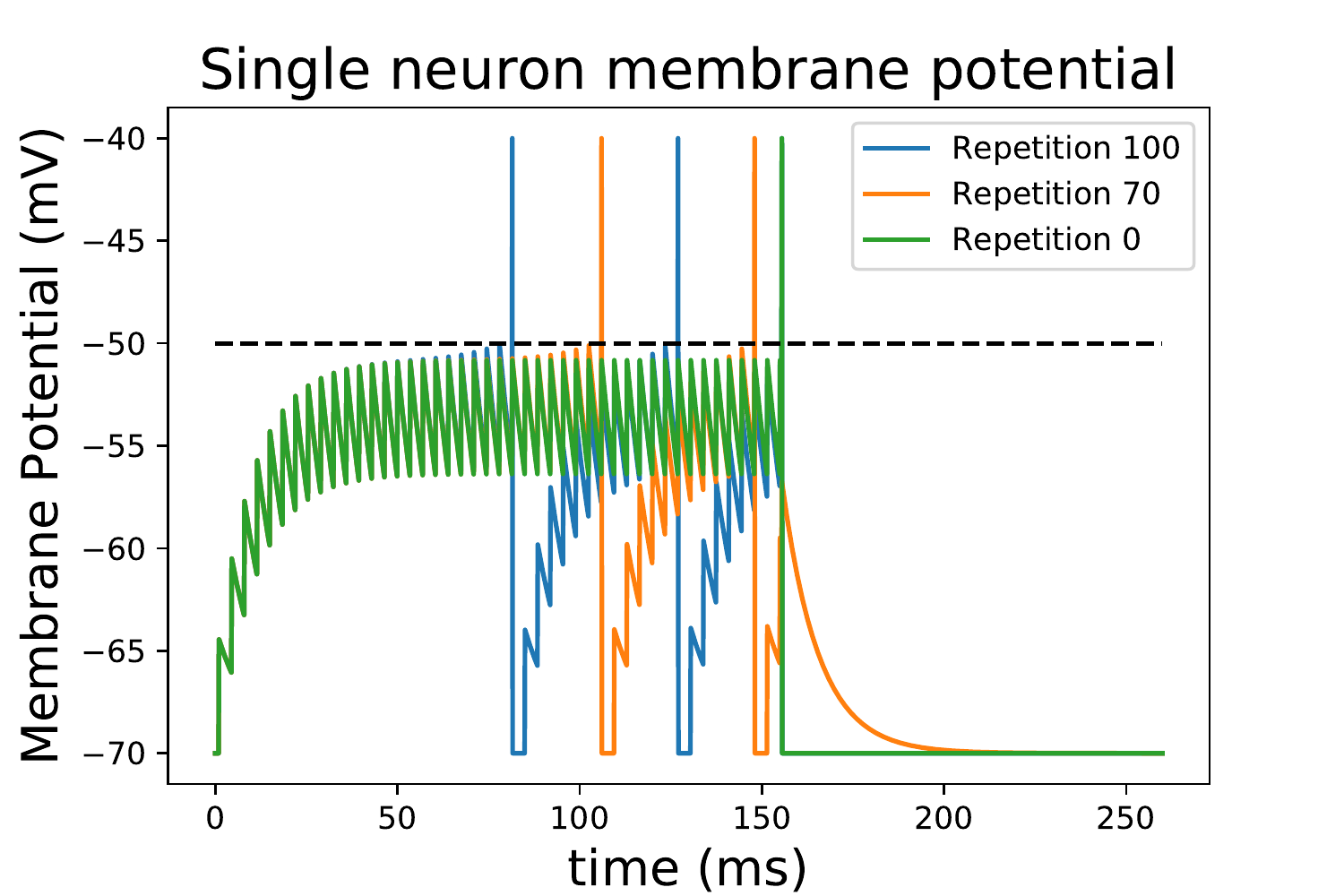}
		\end{subfigure}%
		\begin{subfigure}{0.45\textwidth}
			\centering
			\includegraphics[width=1\linewidth,trim={0cm, 0cm, 1cm, 0cm},clip]{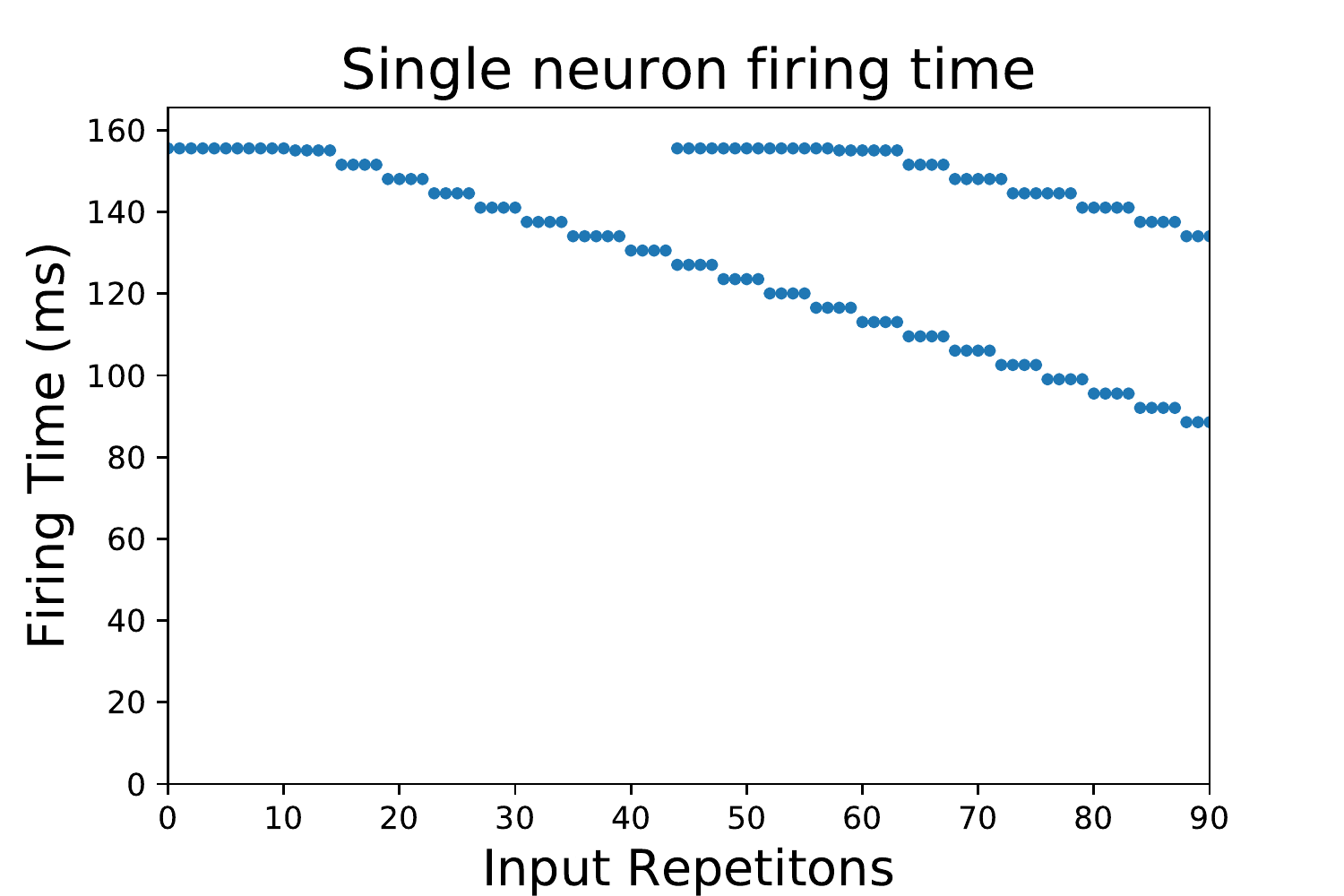}
		\end{subfigure}\\
		\begin{subfigure}{0.48\textwidth}
			\centering
			\includegraphics[width=1\linewidth,trim={0cm, 0cm, 0cm, 0cm},clip]{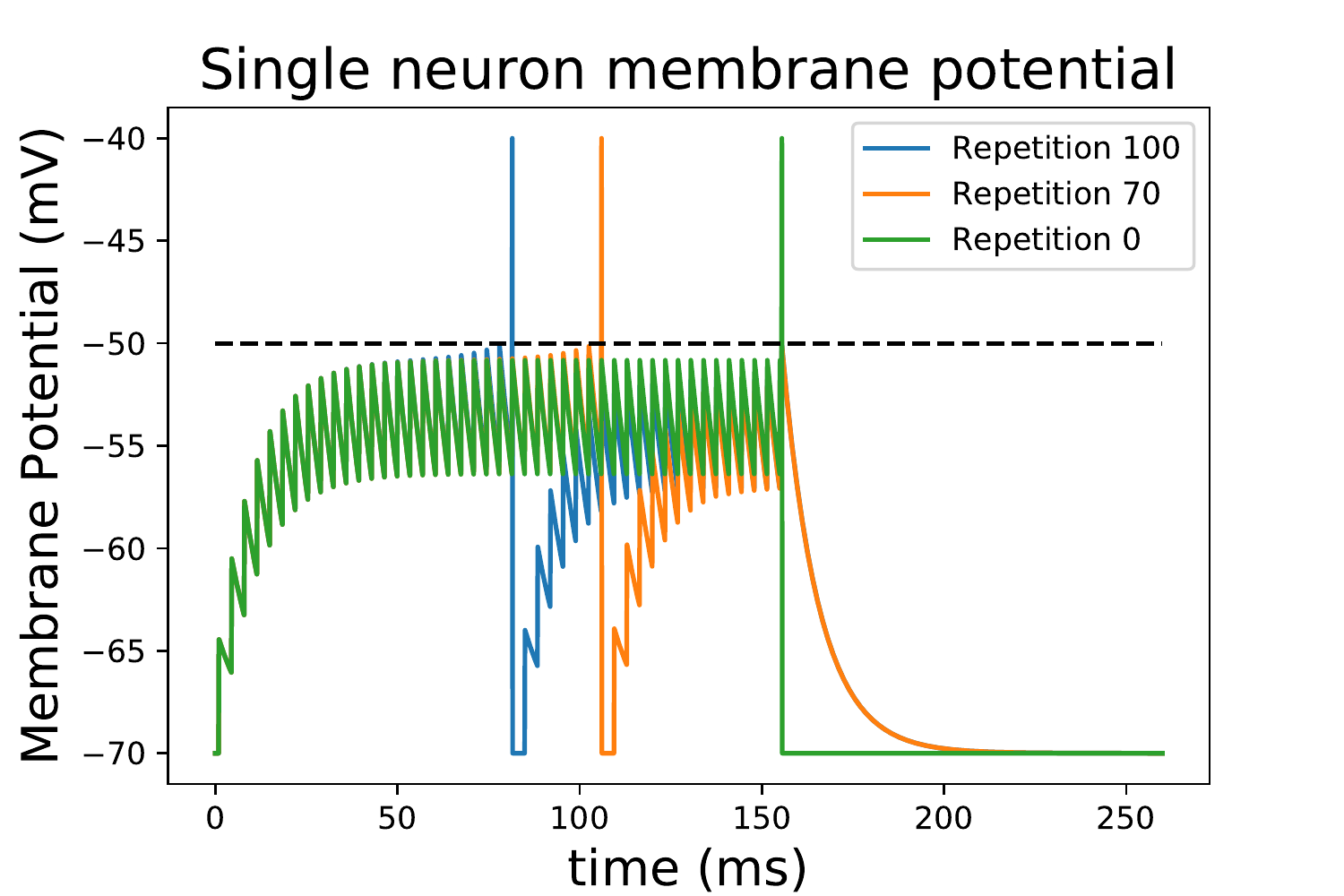}
		\end{subfigure}%
		\begin{subfigure}{0.45\textwidth}
			\centering
			\includegraphics[width=1\linewidth,trim={0cm, 0cm, 1cm, 0cm},clip]{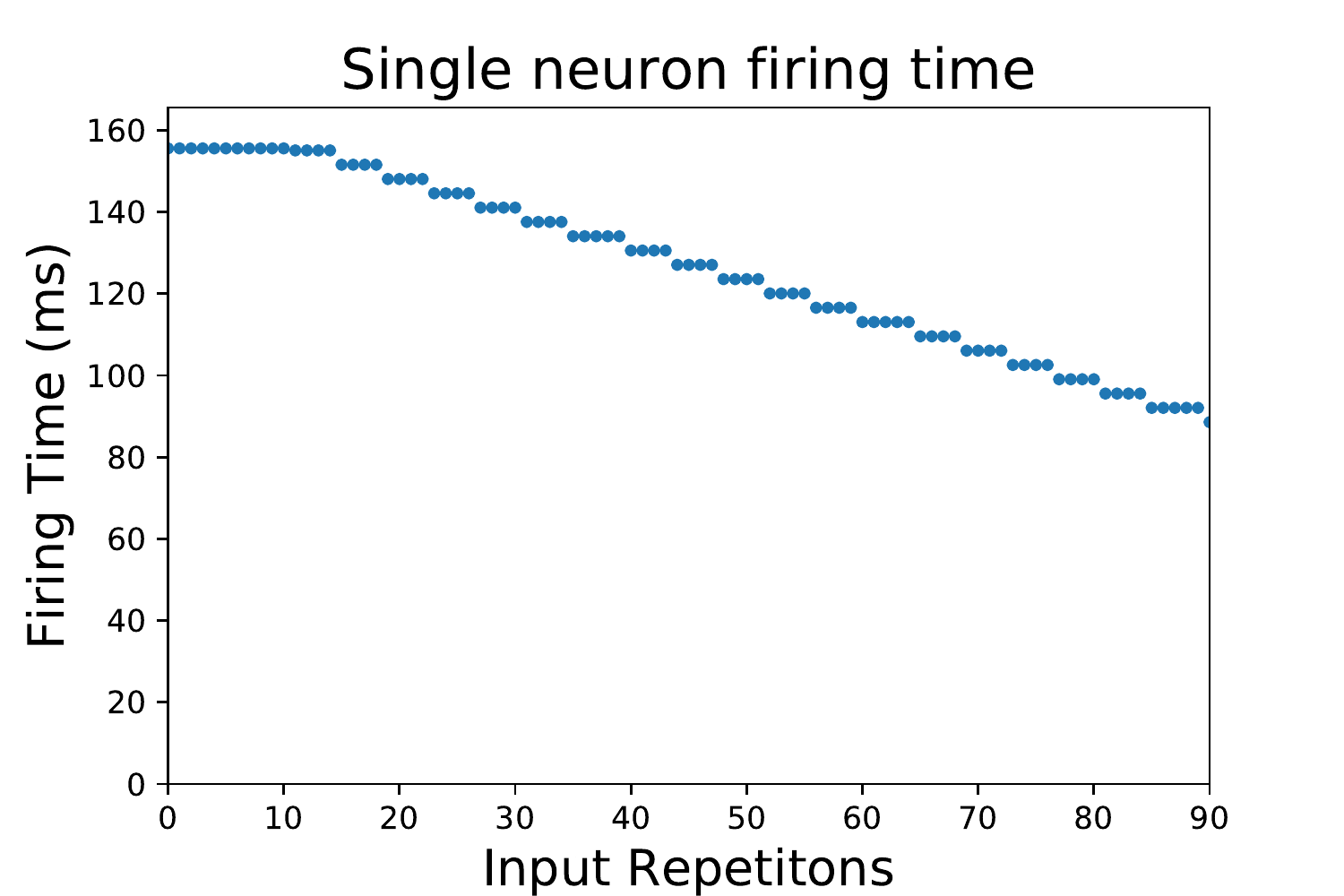}
		\end{subfigure}
		
		\caption{\textbf{Latency reduction and spike proliferation}: We plot the membrane potential (left) and firing times (right) of a postsynaptic neuron that receives a constant train of spikes with inter-spike interval of $3.5ms$ and strength $5.5 mV$, from time $t=0 ms$ to $t=150ms$, and we add an extra spike at $t=150ms$ with potential $2mV$. The neuron generates a single postsynaptic spike at the original input presentation (Repetition 0). The upper plots reflect the case $\eta_+=\eta_-$, while for the lower ones we picked $\frac{3}{2}\eta_+ = \eta_-$. After an initialization period, the postsynaptic spike moves forward in time at a constant rate. As this happens, a single presynaptic spike will get reinforced proportionally to the $\eta_+$ and dampened proportionally to $\eta_-$. If LTP is equal to LTD, after the postsynaptic spike happens much earlier than before, the membrane potential of the postsynaptic neuron will reach the threshold again. This second postsynaptic spike would move forward in time at the same speed as the strengths of the spikes are left unchanged by the compensation of LTD and LTP (upper plots). In the case where $\eta_+<\eta_-$, the depression compensates the potentiation, so there is no second postsynaptic spike. }\label{fig:LTPvsLTD}
	\end{figure}

	This is easy to understand in a simple scenario: Considering a very long, excitatory presynaptic spike train which generates a single postsynaptic spike at some time $t_0$. The postsynaptic spike will advance through the spike train, and after some repetitions it will be triggered one presynaptic spike earlier. After this advancement is repeated many times, the postsynaptic spike is triggered at time $t_\infty$, very far (in time) from the place where it was first triggered, so that 
	\begin{equation}
	t_\infty \ll t_0.
	\end{equation}
	The membrane potential decays exponentially, meaning that the effect of the postsynaptic spike at time $t_\infty$ on the membrane potential are of order $O(e^{-\frac{t_0 - t_\infty }{\tau_m}})$, which is negligible. Thus, the membrane potential at time $t_0$ is now only dependent on the presynaptic spikes that are close. If those presynaptic spikes have been left as they where by the passage of the postsynaptic spike, then a new postsynaptic spike will be generated at time $t_0$. To not have this postsynaptic spike appear, it is therefore necessary that the passage of the postsynaptic spike weakens the presynaptic ones. We illustrate this point in  Fig.~\ref{fig:LTPvsLTD} with the functions and parameters that we will use in subsequent sessions. 
	
	Note that the argument that we give here is qualitative in nature, in the sense that we simply state that LTD should dominate LTP through the constant $\eta$, but we have not studied how to find that ratio. As this would depend on the exact parameters of the regular spike train --and thus would not be directly generalizable--, we will simply assume that the brain operates in a parameter regime in which spikes do not proliferate.

\subsection{Late spike disappearance through synaptic noise}
If latencies might be reduced, then two postsynaptic spikes that are triggered at distant points in time might become close as time progresses. We must then ask what happens to a pair of postsynaptic spikes that occur very close in time. In this section we show that in the absence of synaptic noise the two spikes can coexist, but random modifications of the presynaptic weights --induced, for instance, by other presynaptic inputs -- can lead to the disappearance of the second postsynaptic spike.

There are many possible scenarios that we might consider when we have pairs of postsynaptic spikes in the same neuron: we must consider the time between the two spikes, the movements in time of both of them and the possibility of synaptic noise. The case when two postsynaptic spikes happen originally very close in time is extremely rare -- because postsynaptic spikes are sparse. The case where the first postsynaptic spike also moves is not interesting, because the spike will move forward in time, increasing the distance between the two postsynaptic spikes and thus reducing the LTD effect on the second spike. Therefore we will consider the case where there is an early postsynaptic spike at some fixed time that will remain in place, and a second postsynaptic spike that will initially be triggered very far in time. 

The intuition here is that there is a time interval for the second postsynaptic spike, in which the LTD of the first postsynaptic spike would lead to a decrease in the membrane potential of the postsynaptic neuron at the time of the second postsynaptic spike, which could lead to the irreversible disappearance of the second postsynaptic spike or its recession. Outside of this time interval, the second postsynaptic spike will reduce its latency, approaching the early postsynaptic spike and the dangerous zone. In the remaining of this section we will show that this interval is never reached in a deterministic system but that the addition of noise can enforce this disappearance.

We start by showing that repeating always the same input spike train without noise cannot lead to the reduction of the number of postsynaptic spikes. Consider a long presynaptic spike train with presynaptic spikes arriving at $t_0, t_1,...t_N$, which generates two postsynaptic spikes, one at time $t_0$, which is fixed and will appear at every presentation of the spike train, and another one that is originally triggered at $t_N$. For the second spike to disappear, it can either do so at $t_N$ or first advance through the spike train -- that means, being triggered at $t_{N-1}$, then at $t_{N-2}$ and so on -- and eventually die. For now, we assume that $t_N - t_0 \ll \tau_s$, so that initially the spike at time $t_N$ evolves independently of the spike at time $t_0$, and it would not disappear at $t_N$. Consider now that the input has been repeated long enough so that the second postsynaptic spike is now triggered at $t_i$, and the effects of the STDP generated by the spike at $t_0$ are not negligible to the presynaptic weight $t_{i-1}$, which is associated to the presynaptic spike at $t_{i-1}$. If the postsynaptic spike is originally triggered at $t_i$, then it would move to $t_{i-1}$ only if, after repeating the same input many times,
\begin{equation}
	v(t_{i-1}) = \sum_{k=1}^{i-1} w_k e^{-\frac{t_k-t_{i-1}}{\tau_m}} \geq v_{th}.
\end{equation}
After $v(t_{i-1})$ crosses the $v_{th}$ threshold, the postsynaptic spike at $t_i$ moves to $t_{i-1}$, and thus the time difference between every presynaptic spike at $t\leq t_{i-1}$ and the postsynaptic spike is reduced. This naturally implies that the synaptic weights $w_k$ for all $k\leq i-1$ increase, thus the postsynaptic spike cannot disappear because the membrane potential at $v(t_{i-1})$ can not decrease unless the postsynaptic spike moves to $t_{i-2}$.

This argument assumes that presynaptic spike trains are always repeated with fixed spike timings but with weights that are affected by LTP and LTD. This is generally not true, as there are many factors that can introduce stochasticity on the evolution of the weights, such as jitter, the stochastic nature of molecular dynamics on the synaptic cleft and on the neuron membrane. 

If we now consider the stability of both postsynaptic spikes with respect to that noise, we easily realize that they are not equal: while the presynaptic spikes that generate the first postsynaptic spike are only subject to LTP and noise, the presynaptic spikes that generate the second spike -- which happen necessarily between postsynaptic spikes -- are subject to both LTP -- from the late postsynaptic spike -- and LTD --from the earlier postsynaptic spike -- on top of the noise. 

This difference implies that the noise can make a postsynaptic spike disappear or recede, either by directly weakening the associated presynaptic weights or strengthening them, so that the postsynaptic spike moves into a region where LTD dominates and it would be later erased or receded.

To explain this in the setting that we used before, consider a neuron with a postsynaptic spike at time $t_i$ that would not move to $t_{i-1}$ in the previous deterministic system. However, now the weights evolve by the combined effects of that spike, an earlier postsynaptic spike at time $t_0$ and some noise. The membrane potential at time $t_i$ and after $r$ repetitions of the input spike train follows
\begin{equation}
	v(t_{i}) = \sum_{k=1}^{i} w_k e^{-\frac{t_k-t_{i}}{\tau_m}} + \xi_{t_{i}},
\end{equation}
where $\xi_t$ is the contribution of the random evolution of the weights to $v(t)$ given by
\begin{equation}
	\xi_{t_i} = \sum_{k=1}^i \delta w_k e^{-\frac{t_k-t_{i}}{\tau_m}}
\end{equation}
where $\delta w_k$ is the deviation of weight $w_k$ from its deterministic evolution; in the case of Gaussian noise, for instance, it would lead to an Ornstein--Uhlenbeck process for the evolution of $v(t_i,r)$ on the repetition variable $r$. Note that the noise is not a variable reinitialized at every repetition, as it has some momentum as weights evolve slowly.

\begin{figure}[h]
 	\begin{subfigure}{0.48\textwidth}
 		\centering
 		\includegraphics[width=1\linewidth,trim={0cm, 0cm, 0cm, 0cm},clip]{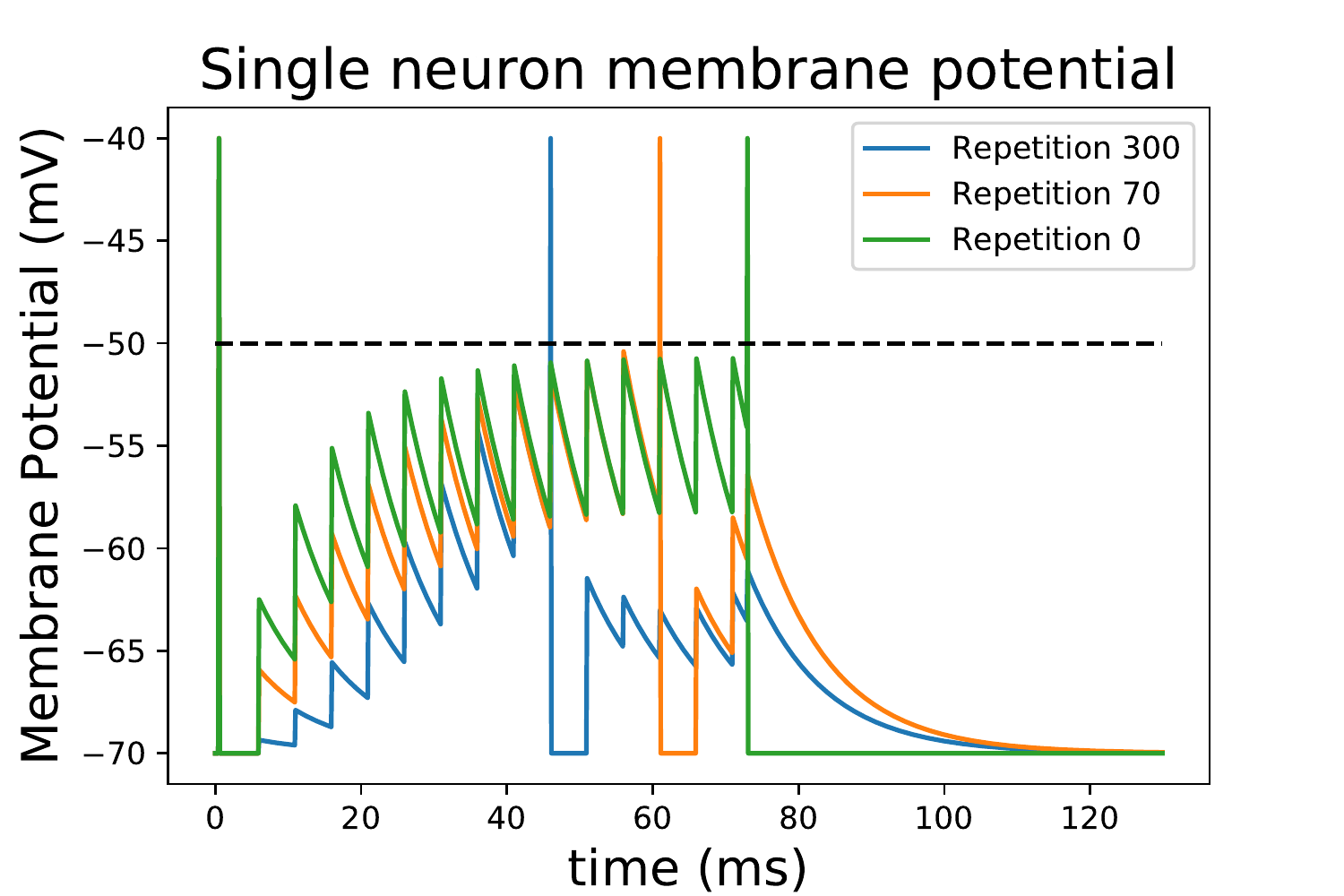}
 	\end{subfigure}%
 	\begin{subfigure}{0.45\textwidth}
 		\centering
 		\includegraphics[width=1\linewidth,trim={0cm, 0cm, 1cm, 0cm},clip]{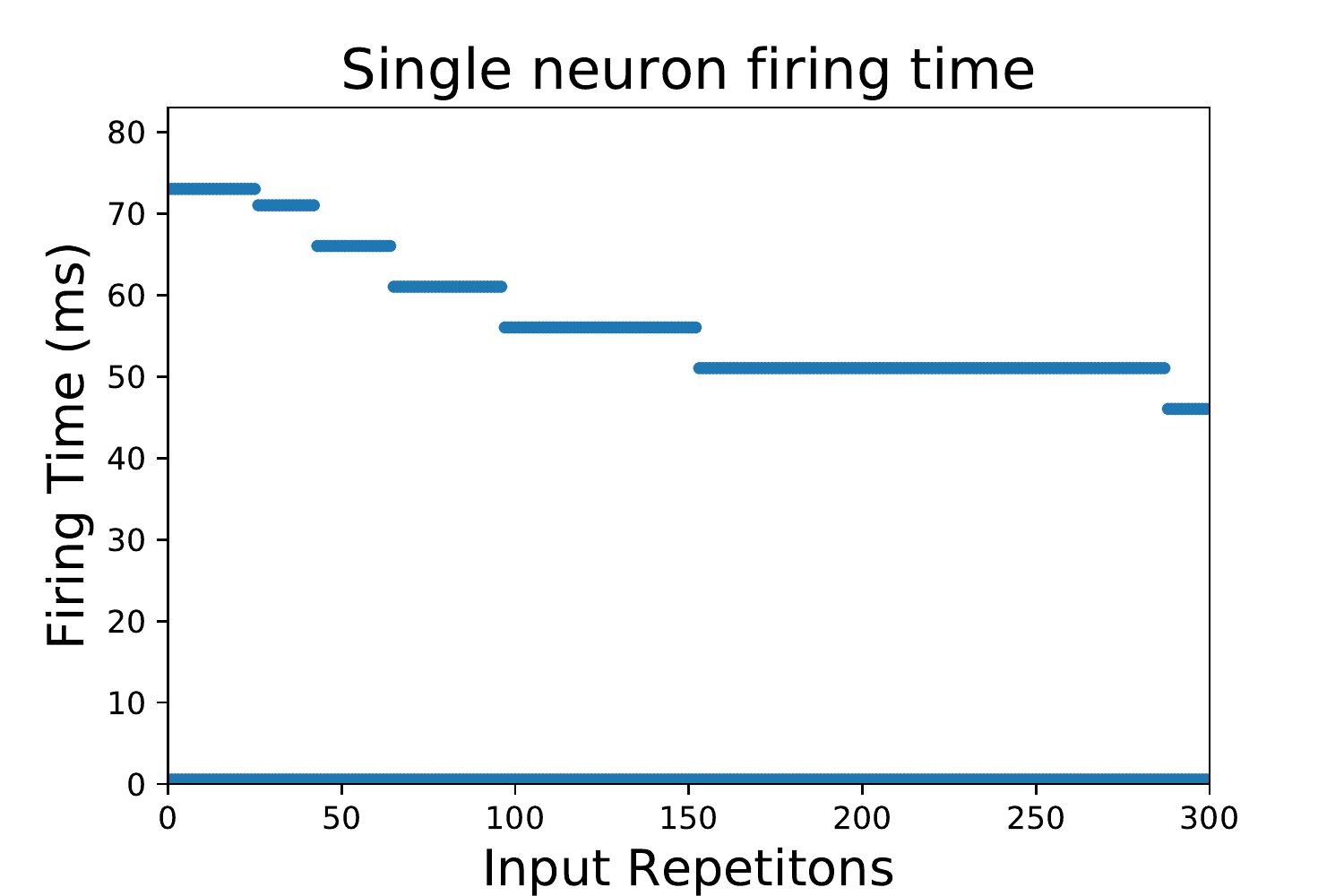}
 	\end{subfigure}\\
 	\begin{subfigure}{0.48\textwidth}
 		\centering
 		\includegraphics[width=1\linewidth,trim={0cm, 0cm, 0cm, 0cm},clip]{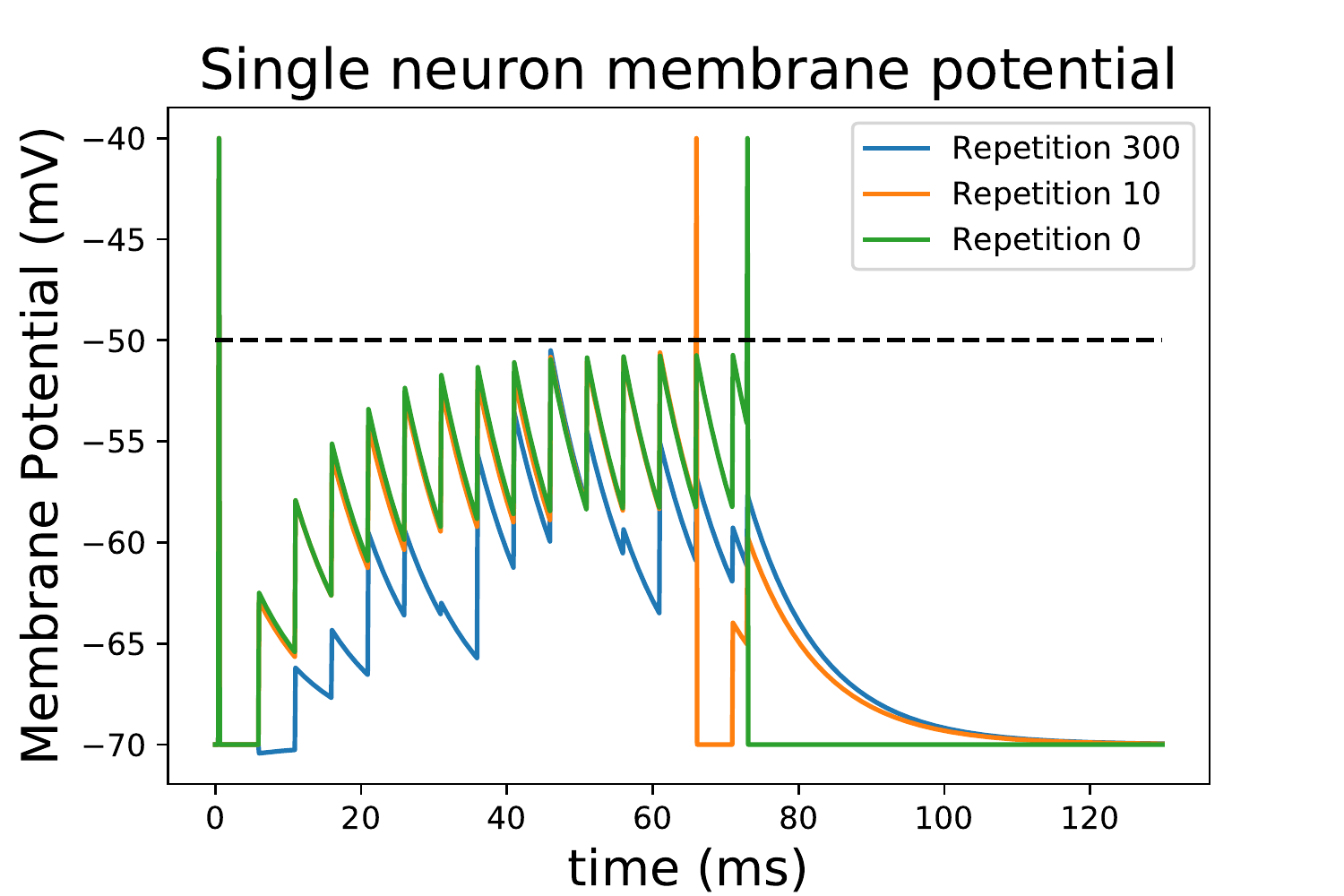}
 	\end{subfigure}%
 	\begin{subfigure}{0.45\textwidth}
 		\centering
 		\includegraphics[width=1\linewidth,trim={0cm, 0cm, 1cm, 0cm},clip]{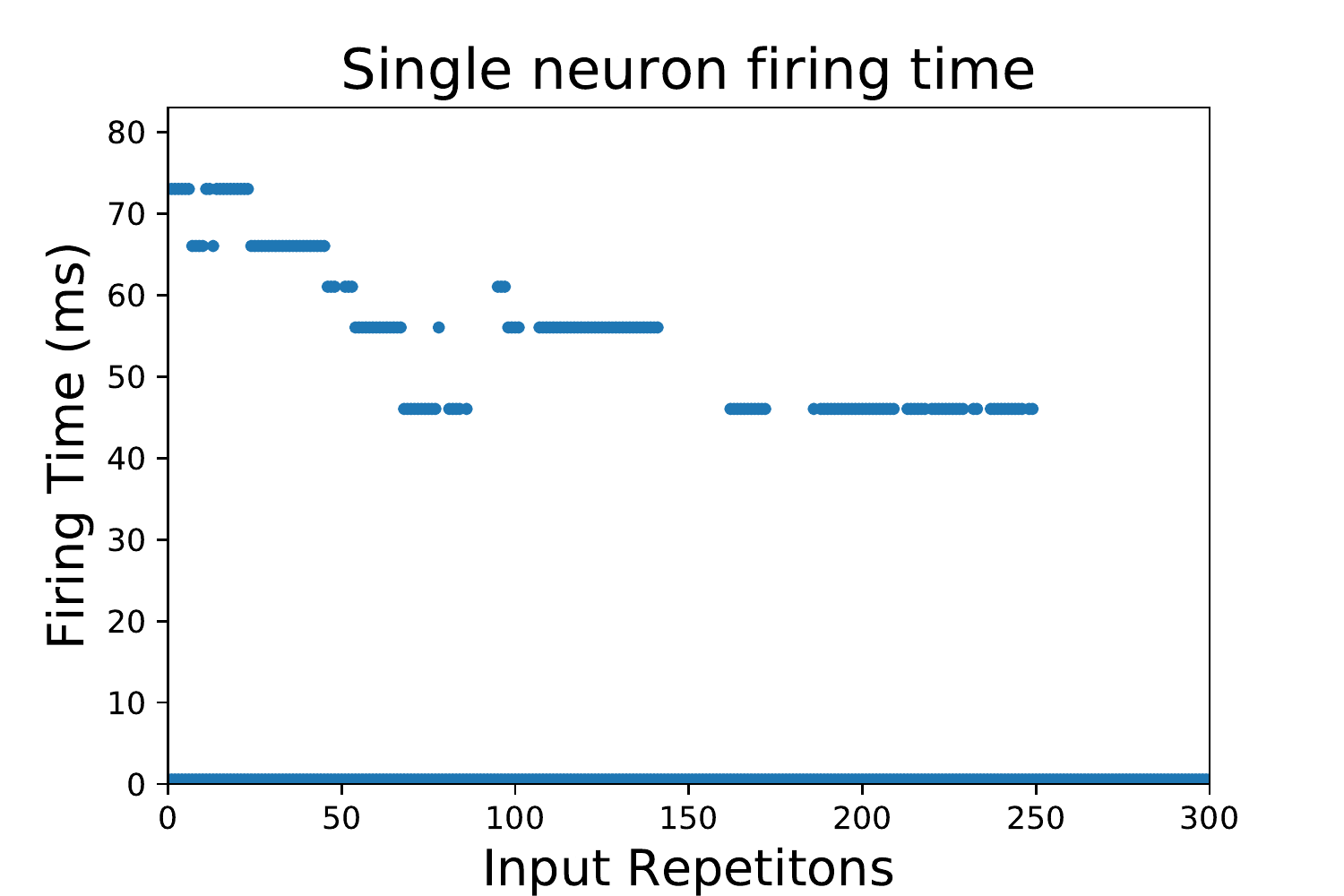}
 	\end{subfigure}
 	\caption{\textbf{Noise deletes late spike in a regular presynaptic spike train}: We plot the membrane potential (left) and firing times (right) of a postsynaptic neuron that receives a constant train of spikes with inter-spike interval of $5ms$ and strength $7.5 mV$, from time $t=0 ms$ to $t=150ms$, and we add an extra spike at $t=80ms$ with potential $5mV$, with a forces postsynaptic spike at time $0.5ms$. Note also that, during its existence, the latency of the postsynaptic spike subject to noise decreases faster than its noiseless counterpart. }\label{fig:RegReductionNumberSpikes}
 \end{figure}

If this postsynaptic spike train is repeated very often, the deterministic part of the weights goes to a fixed value, which is zero for $k>i$ and thus $v(t_k)\sim \xi_{t_k}$ for all $k>i$. Thus, under the assumption 
\begin{equation}
	\xi_{t_{i}} < \sum_{k=1}^{i} w_k e^{-\frac{t_k-t_{i}}{\tau_m}} - v_{th}
\end{equation}
for a few repetitions of the input spike train, the seconds spike vanishes. Therefore among the subsequent input repetitions subjected to the ever present postsynaptic spike at $t_0, $the weights $w_k$ will decrease for all values of $k$, hence it is possible that $v(t_i) < v_{th}$ thereafter. This will result in the irreversible disappearance of the postsynaptic spike at $t_i$ or its delay. This is illustrated in Fig.~\ref{fig:RegReductionNumberSpikes}.

\subsection{Generalization to inhibitory plasticity}\label{sec:inhibitoryPlasticity}
Until now we have considered only excitatory neurons. However, in biological systems, inhibitory synapses are also present and have plasticity \cite{vogels2013inhibitory}. Naturally, this might compromise the effects described in the previous section, as an inhibitory synapse that gets potentiated could counteract the effects of excitatory STDP. For instance, it might decrease the membrane potential and thus increase the latency of the postsynaptic neuron \cite{effenberger2015self}. Our goal in this section is to find the parameter regime in which the presence of inhibitory plasticity does not compromise the latency decrease and, by extension, the disappearance of postsynaptic spikes.

Intuitively, as long as the STDP in inhibitory synapses is weaker than the STDP in excitatory ones, the latency of postsynaptic spikes would still be decreased. The question is then to find a way of measuring "how much weaker" it has to be. To address this issue, we must find a boundary parameter set for inhibitory synapses that guarantees that latency would be reduced, and then we can simply take any parameter set that is between this boundary parameter set and the only-excitatory STDP.

To identify the parameter regime in which latency reduction for a single spike appears, we assume that the STDP keeps the balance between excitation and inhibition, in the sense  that the average input to a single neuron is maintained \cite{brunel2000dynamics}. To maintain this balance, the potentiation of excitatory synapses is compensated by the potentiation of inhibitory synapses. Potentiating all synapses but maintaining the average input leads to the increase in fluctuations of the membrane potential, meaning that the membrane potential preceding a postsynaptic spike would change more around the average, and thus it can still lead to an earlier postsynaptic spike.

Consider a single postsynaptic spike at time $t_{pos}$. For $t<t_{post}$,
\begin{equation}
v(t) = \sum_{t_k<t} w_k e^{-\frac{t-t_k}{\tau_m}},
\end{equation}
and initially $v(t) < v_{th}$. Now we wonder what happens when the weights $w_k$ change, specifically if the postsynaptic spike would advance, recede or disappear. This depends on the exact values of $w_k$ and $t_k$, so to make more generic statements we are interested in the value
\begin{equation}
\mathbb{E}_r\left[\Delta t_{post}\right] = \mathbb{E}\left[ t^r_{post} - t_{post}\right]
= \left(\mathbb{E}\left[ t^r_{post}\right] -t_{post}\right)\text{Pr}\left[\exists s\right] 
\end{equation}
where $r$ accounts for the number of times that the spike train has been repeated, and $\text{Pr}\left[\exists s\right]$ is the probability that a postsynaptic spike still exists, and the expectations are taken over the presynaptic spike trains -- a list of tuples $(w_k,t_k)$ sampled from some predefined distribution-- that generate a postsynaptic spike at time $t_{post}$. In simpler words, we are simply trying to calculate whether the postsynaptic spike is expected to move forward ($\mathbb{E}\left[\Delta t_{post}\right](r)< 0$) or backward ($\mathbb{E}\left[\Delta t_{post}\right](r)< 0$), ignoring the ones that disappeared, if we only have some information about the distribution from which the list of  $(w_k,t_k)$ was sampled.

We know that increasing the input excitatory weights can only lead to an earlier postsynaptic spike, because $v(t)$ can only increase and thus it might reach $v_{th}$ earlier. We will take this a step further and assume that this statement is also true about the average weights, meaning that when the expected input increases, the expected postsynaptic firing time also increases. In more formal terms, we are assuming that $\mathbb{E}_r\left[\Delta t_{post}\right]$ is a function that decreases monotonically with 
\begin{equation}
\mathbb{E}\left[\Delta_r v(t)\right] 
= \mathbb{E}\left[\int_{-\infty}^{t}\Delta_r i(t)e^{-\frac{t-x}{\tau_m}}dx\right] 
= \int_{-\infty}^{t}\mathbb{E}\left[\Delta_r i(t)\right]e^{-\frac{t-x}{\tau_m}} dx,
\end{equation}
for all $t<t_{post}$, meaning that if the expected value of $\Delta v(t)$ averaged over all realizations of the input spike train producing a spike at $t_{post}$ is positive, then $\mathbb{E}_r\left[\Delta t_{post}\right]$ will be negative. 

This assumption, albeit natural, requires some careful consideration. Specifically, we must clarify the distribution over which the expectations are taken, which corresponds to all possible presynaptic spike trains shortly preceding a postsynaptic spike. Those spike trains have fixed timings for every postsynaptic spike under consideration, but are updated systematically because postsynaptic spikes evolve with the input repetitions and the noise. Thus, this distribution considers samples in which a new spike has just appeared or samples where a postsynaptic spike has recently been displaced by a short time.

The subsequent step is to find the conditions that guarantee that $\mathbb{E}\left[\Delta_r v(t)\right] $ increases. A sufficient condition for this to happen is to have 
\begin{equation}\label{eq:CondIneqIncreaseInput}
\mathbb{E}\left[\Delta_r i(t)\right] = \Delta_r \mathbb{E}\left[i_e(t)\right]- \Delta_r\mathbb{E}\left[i_i(t)\right]> 0,\ \forall t< t_{post}
\end{equation}
where $\mathbb{E}\left[i_e(t)\right]$ is the expected input to the neuron at time $t$, and $ \mathbb{E}\left[i_i(t)\right]$, $\mathbb{E}\left[i_i(t)\right]$ is simply its decomposition in inhibitory and excitatory inputs, which gives us
\begin{equation}
\begin{aligned}
\mathbb{E}\left[i_e(t)\right] = \rho_e\int_{0}^{\infty} \mu_{w_e}(w,t) dw\\
\mathbb{E}\left[i_i(t)\right] = \rho_i\int_{0}^{\infty} \mu_{w_i}(w,t) dw
\end{aligned}
\end{equation}
where $\rho_e, \rho_i$ are the rates of incoming spikes and $\mu_{w_e}(w,t), \mu_{w_i}(w,t)$ the probabilities of the weights associated to time $t$. 

Thus, to maintain the condition from Eq.~\ref{eq:CondIneqIncreaseInput} we must ensure that the parameters $\mu_{w_e}$, $\mu_{w_i}$, $\eta_+^e$, $\eta_+^i$, $w_{\min}^e$, $w_{\min}^i$ are such that
\begin{equation}
\rho_e\int_{0}^{\infty} \Delta w_e(r) \mu_{w_e}(w,t) dw > \rho_i\int_{0}^{\infty} \Delta w_i(r)\mu_{w_i}(w,t) dw,
\end{equation}
where $\Delta w(r)$ are given by the STDP Eq.~\ref{eq:STDP_1} over many repetitions --counted by $r$ -- of the input spike train. We will now find a parameter regime in which this holds by finding its boundary. In other words, we are interested on the parameter set in which
\begin{equation}
\rho_e\int_{0}^{\infty} \Delta w_e(r) \mu_{w_e}(w,t) dw = \rho_i\int_{0}^{\infty} \Delta w_i(r)\mu_{w_i}(w,t) dw.
\end{equation}
Note that it is not enough to find two weight distributions $\mu_{w_e}$, $\mu_{w_i}$ where
\begin{equation}
\rho_e\int_{0}^{\infty} A_+^e(w_e) \mu_{w_e}(w,t) dw = \rho_i\int_{0}^{\infty} A_+^i(w_i)\mu_{w_i}(w,t) dw,
\end{equation} 
because this would only work for the first input repetition. We have to ensure that even after STDP changes the distribution, the equality holds. A simple way to achieve this is to set the inhibitory and excitatory parameters to be equal. It is obvious that if the probability distributions of weights, input rates and STDP parameters are the same, then the change in input will affect the inhibitory and excitatory synapses in the same way. However, we know that this is not the case, as there are typically fewer inhibitory synapses than excitatory ones. Thus, we modify this symmetry to include rescaling, meaning that we have the ratio
\begin{equation}
\alpha = \dfrac{\rho_i}{\rho_e}
\end{equation}
that is also intrinsic to the probability distributions
\begin{equation}\label{eq:ExInBalanceCond1}
\alpha \mu_{w_i}\left(\alpha x,t\right) = \mu_{w_e}(x,t) \ \forall x,t.
\end{equation}
and the STDP parameters
\begin{equation}\label{eq:ExInBalanceCond2}
\alpha A_+^i(\alpha x) = A_+^e(x)\ \forall x.
\end{equation}
By a simple change of variable we can show that, if those properties are satisfied,
\begin{equation}
\begin{aligned}
\rho_e\int_{0}^{\infty} &A_+^e(x) \mu_{w_e}(x,t) dx =
\dfrac{1}{\alpha} \rho_i \int_{0}^{\infty} \alpha A_+^i(\alpha x) \alpha \mu_{w_i}\left(\alpha x,t\right)  \frac{1}{\alpha} d(\alpha x)\\
&=
\rho_i\int_{0}^{\infty} A_+^i(y)\mu_{w_i}(y,t) dy.
\end{aligned}
\end{equation}
Furthermore, if we take a pair of inhibitory and excitatory weights such that $w_e = \alpha w_i$ we have that after STDP,
\begin{equation}
\alpha w_i \rightarrow \alpha (w_i + A_+^i(w_i)) = \alpha w_i + \alpha A_+^i(\alpha w_e) = w_e + A_+^e( w_e) \leftarrow  w_e,
\end{equation}
meaning that the weight probability changes in such a way that
\begin{equation}
\mu_{w_e}'\left(x + A_+^e(x),t\right) = \mu_{w_e}\left(x,t\right)  = \alpha \mu_{w_i}\left(\alpha x,t\right) = \alpha \mu_{w_i}'\left(\alpha \left(x + A_+^i(x)\right),t\right),
\end{equation}
where $\mu_{w_e}'$ and $\mu_{w_i}'$ are the weight distributions after STDP has acted once. Thus, if Eq.~\ref{eq:ExInBalanceCond1} holds at some point, it will also hold for all subsequent iterations of the input spike pattern.

Thus, we have found a set of conditions that satisfy Eq.~\ref{eq:ExInBalanceCond1} at $r=0$ and for any subsequent $r>0$ for the case where the postsynaptic spike does not change during the $r$ repetitions. Notice that the self-consistency of this condition does not make any assumptions about the learning constant or $\Delta t$ dependent term on STDP, or even its sign, it only requires that the expected increase --or decrease --  in excitatory input is matched by the expected increase -- or decrease -- in inhibitory input. In particular, this symmetry does not change if the postsynaptic spike advances places, because the STDP kernel has the same ratio of potentiating inhibitory and excitatory synapses. In other words, when a postsynaptic spike changes places before the $r$th input repetitions, the variance of the input before the postsynaptic spike still increases and, converselly, it still decreases after the postsynaptic spike. 


Now we have a large set of parameters in which latency reduction is expected to happen. Any STDP parameters for which $\alpha A_+^i(\alpha x) <  A_+^e( x)$ combined with Eq.~\ref{eq:ExInBalanceCond1}, or distribution of weights with  $\alpha \mu_{w_i}\left(\alpha x,t\right) < \mu_{w_e}(x,t)$ with Eq.~\ref{eq:ExInBalanceCond2}, or both cases combined.

It is worth noticing that the case when all the equalities Eq.~\ref{eq:ExInBalanceCond1} and Eq.~\ref{eq:ExInBalanceCond2} are met we would still expect the latency to decrease. The reason is that even if 
\begin{equation}
\mathbb{E}\left[\Delta_r v(t)\right]  = \mathbb{E}\left[\Delta_r i(t)\right]  = 0,
\end{equation}
the variance of $v(t)$ increases. More explicitly, 
\begin{equation}
\begin{aligned}
\Delta_r\text{Var}&\left[v(t)\right] 
= \Delta_r \int_{-\infty}^{t}\text{Var}\left[i(t)\right] dt  
= \Delta_r \int_{-\infty}^{t}\left(\mathbb{E}\left[i^2(t)\right] - \mathbb{E}\left[i(t)\right]^2\right)dt   \\
= \Delta_r& \int_{-\infty}^{t}\mathbb{E}\left[i^2(t)\right] dt = 
\int_{-\infty}^{t} \Delta_r \mathbb{E}\left[i_e^2(t)\right]dt
+ \int_{-\infty}^{t} \Delta_r \mathbb{E}\left[i_i^2(t)\right]dt
\end{aligned}
\end{equation}
where the term $\mathbb{E}\left[i(t)\right]^2 = 0$ by the symmetry of the weights and it is maintained at zero by the symmetry of the STDP. Since we are only concerned with $t<t_{post}$, STDP potentiates both inhibitory and excitatory synapses, so 
\begin{equation}\label{eq:covarInputs}
\begin{aligned}
	\Delta_r \mathbb{E}\left[i_i^2(t)\right] ,\Delta_r \mathbb{E}\left[i_e^2(t)\right] > 0
\end{aligned}
\end{equation}
and therefore the variance increases.

Naturally, if the variance of a certain distribution increases, then the probability of reaching a value higher than some threshold --$v_{th}$-- also increases. Notice that even with an increase on variance of $v(t)$ we would not expect the postsynaptic spike to recede, as the inputs $i(t)$ at $t\lesssim t_{post}$ are excitatory --because that $v(t_{post})>v_{th}$-- and thus by Eq.~\ref{eq:covarInputs}, the membrane potential at $v(t_{post})$ should increase.

The approach outlined here can be also used for other STDP kernels. While the symmetry in the excitatory and inhibitory STDP kernels might not exist for some choices of inhibitory and excitatory plasticity, the approach can, in principle, still be used by simply adding a bias on the number or weight of excitatory synapses that would compensate the asymmetry in the STDP kernel. In the interest of simplicity we will also skip this.

The non-proliferation of spikes can be derived by a similar argument, although in this case the mean or the variance (or both) of the presynaptic input to the postsynaptic neuron will decrease due to the depressive nature of STDP for $t>t_{post}$. In general, the idea of having the depression stronger than the potentiation would still work, as long as the depression of inhibitory synapses is weaker or equal than that of excitatory synapses. As this calculation is essentially the same as the one we just presented, we will omit it.

\subsection{Numerical verification for random input spike trains}

The examples presented to illustrate the latency reduction and the disappearance of late postsynaptic spikes were simple, so we must now extend them to a more general case. To do so, we simulated spike trains where the times of the presynaptic spikes are random, including only excitatory or excitatory and inhibitory STDP, noise and the presence of an earlier postsynaptic spike. The results are presented in Table~\ref{tab:ShortTimescales} and agree with our previous conclusions: a single postsynaptic spike tends to reduce its latency, if there are multiple postsynaptic spikes in a short time window the later one tends to disappear, and the presence of noise increases those effects. Note that we have not included jitter or probabilistic presynaptic spikes, choosing instead to have noise directly on the weight evolutions. As both cases have been addressed before \cite{guyonneau2005neurons} with similar conclusions, we shall not repeat them here.

\begin{table}[]
	\small
	\begin{tabular}{|l|l|l|l|l|l|l|l|}
		\hline
		\textbf{\begin{tabular}[c]{@{}l@{}}STDP\\ Type\end{tabular}} & \textbf{\begin{tabular}[c]{@{}l@{}}Noise\\ Var.\end{tabular}} & \textbf{\begin{tabular}[c]{@{}l@{}}Spike \\ at t=0\end{tabular}} & \textbf{\begin{tabular}[c]{@{}l@{}}Spike\\ Count\\ Increase\end{tabular}} & \textbf{\begin{tabular}[c]{@{}l@{}}Spike\\ Count\\ Decrease\end{tabular}} & \textbf{\begin{tabular}[c]{@{}l@{}}Spike\\ Latency\\ Increase\end{tabular}} & \textbf{\begin{tabular}[c]{@{}l@{}}Spike\\ Latency\\ Decrease\end{tabular}} & \textbf{\begin{tabular}[c]{@{}l@{}}Average\\ Latency\\ Change\end{tabular}} \\ \hline
		E \& I & 0 & No & 0.04 \% & 11.6 \% & 6.0 \% & 19.7 \% & -1.33 ms \\ \hline
		E \& I & 0.2 & No & 0.2 \% & 23.6 \% & 16.2 \% & 14.8 \% & -1.58 ms \\ \hline
		E & 0 & No & 1.1 \% & 0.0 \% & 1.0 \% & 46.6 \% & -1.27 ms \\ \hline
		E & 0.2 & No & 4.7 \% & 4.1 \% & 7.5 \% & 39.5 \% & -0.89 ms \\ \hline
		E \& I & 0 & Yes & 0.0 \% & 11.3 \% & 6.17 \% & 20.1 \% & -1.27 ms \\ \hline
		E \& I & 0.2 & Yes & 0.2 \% & 24.6 \% & 16.4 \% & 14.7 \% & -1.65 ms \\ \hline
		E & 0  & Yes & 2.8 \% & 4.1 \% & 7.5 \% & 39.7 \% & -0.89 ms \\ \hline
		E & 0.2 & Yes & 2.8 \% & 3.96 \% & 7.2 \% & 39.8 \% & -0.91 ms\\ \hline
	\end{tabular}
	\caption{\textbf{Effects of STDP on short random spike trains}: We explored the effects of STDP on the postsynaptic spike train of a neuron receiving $8$ excitatory and $2$ inhibitory presynaptic spikes arriving at uniformly sampled times on the interval $\left[0, 30 ms\right]$ and the stimulus is repeated $20$ times. The first three columns determine the set-up: the STDP Type indicates if STDP was active for excitatory presynaptic neurons only (E) or for inhibitory as well as excitatory (E \& I) with the inhibitory STDP having the parameters to exactly compensate the excitatory one as presented in Section~\ref{sec:inhibitoryPlasticity}, the second column indicates the variance of the Gaussian noise added to every weight at every stimulus repetition, and the third column indicates whether we added a postsynaptic spike at the begining of the time window. The remaining columns explain the results: the fourth one indicates the percentage of spike trains in which new postsynaptic spikes appeared, the fifth one the percentage of spike trains in which a spike disappeared, the sixth one the percentage of spike trains in which a single postsynaptic spike (not counting the imposed one at $t=0$) happened later after learning, the seventh one corresponds to the postsynaptic spike happening earlier, and the last one is the average latency change of the postsynaptic spikes (here we only accounted for the cases where there was a single postsynaptic spike at the beginning and at the end of the learning). We calculated the percentages and averages from $10000$ randomly generated spike trains in which a single postsynaptic spike was triggered at the beginning of the training. The results clearly show that in all cases spike latencies tend to decrease as well as the number of spikes. Naturally, if STDP affects only excitatory synapses and there is no noise, spikes do not disappear nor do they happen later. Also, as expected adding a postsynaptic spike at $t=0$ reduces the strength of the subsequent postsynaptic spikes, so they tend to disappear more and have stronger latency increases.}\label{tab:ShortTimescales}
\end{table}

So far we have only considered effects in small time scales, meaning that there were only a few spikes on a time interval of the order of $10 ms$, and the postsynaptic spike train would evolve over a few repetitions, on the order of $20$. This leads us to the conclusion that, with plausible assumptions on the parameters of our model, an individual postsynaptic neuron will fire a given postsynaptic spike earlier after many repetitions of the same presynaptic spike train and that if two postsynaptic spikes are close in time, then the later one could disappear. 

\section{Postsynaptic Spike Train}\label{sec:postSpikeTrain}

Now we study the effects of the previously described phenomena, which act on small temporal scales and affect only one or two postsynaptic spikes, for a population of postsynaptic neurons, each one receiving many presynaptic spike trains happening over time scales much larger than $\tau_m$ or $\tau_s$. Specifically, we will explore the latency reduction and suppression or delaying of late postsynaptic spikes change the postsynaptic spike distribution.

Before studying the previous effects, we must validate some of the assumptions that we made in the previous section. Specifically, we assume that all the input spikes came from different synapses, which allowed us to treat the weights of all presynaptic spike as independent. This is a valid assumption when we are considering a short time intervals, as the sparsity of presynaptic firing and the existence of refractory periods implies that a single synapse would typically not fire more than once during a short presynaptic spike train. However, when there is a long presynaptic spike train, a presynaptic neuron might contribute to that spike train more than once, thus our assumption might be invalid and the phenomena described in the previous section might not appear. To ensure that the phenomena of latency reduction and late spike disappearance are still present in long spike trains we use a combinatorial argument and count the number of synapses that might evolve in a non-trivial fashion, which we present in Appendix \ref{app:uncorrelatedSpikes}.

We can now consider the first time an input presynaptic spike train is presented. Every neuron starts at $v(0) = 0$ and then its membrane potential will change depending on its inputs. As the input spike train consists of independent spikes with independent weights, the times of the first spike have a probability distribution $f_0^{1}(t)$ with support on $t>0$, which depends on the parameters of the input spike train. After spiking every neuron resets its membrane potential to zero, and thus the distribution of inter-spike intervals $f_0^{ISI}(t)$ follows
\begin{equation}
	f_0^{ISI}(t) = f_0^{1}(t-t_{ref}).
\end{equation}

\begin{figure}[h]
	\begin{subfigure}{0.49\textwidth}
		\centering
		\includegraphics[width=1\linewidth,trim={0cm, 0cm, 1cm, 0cm},clip]{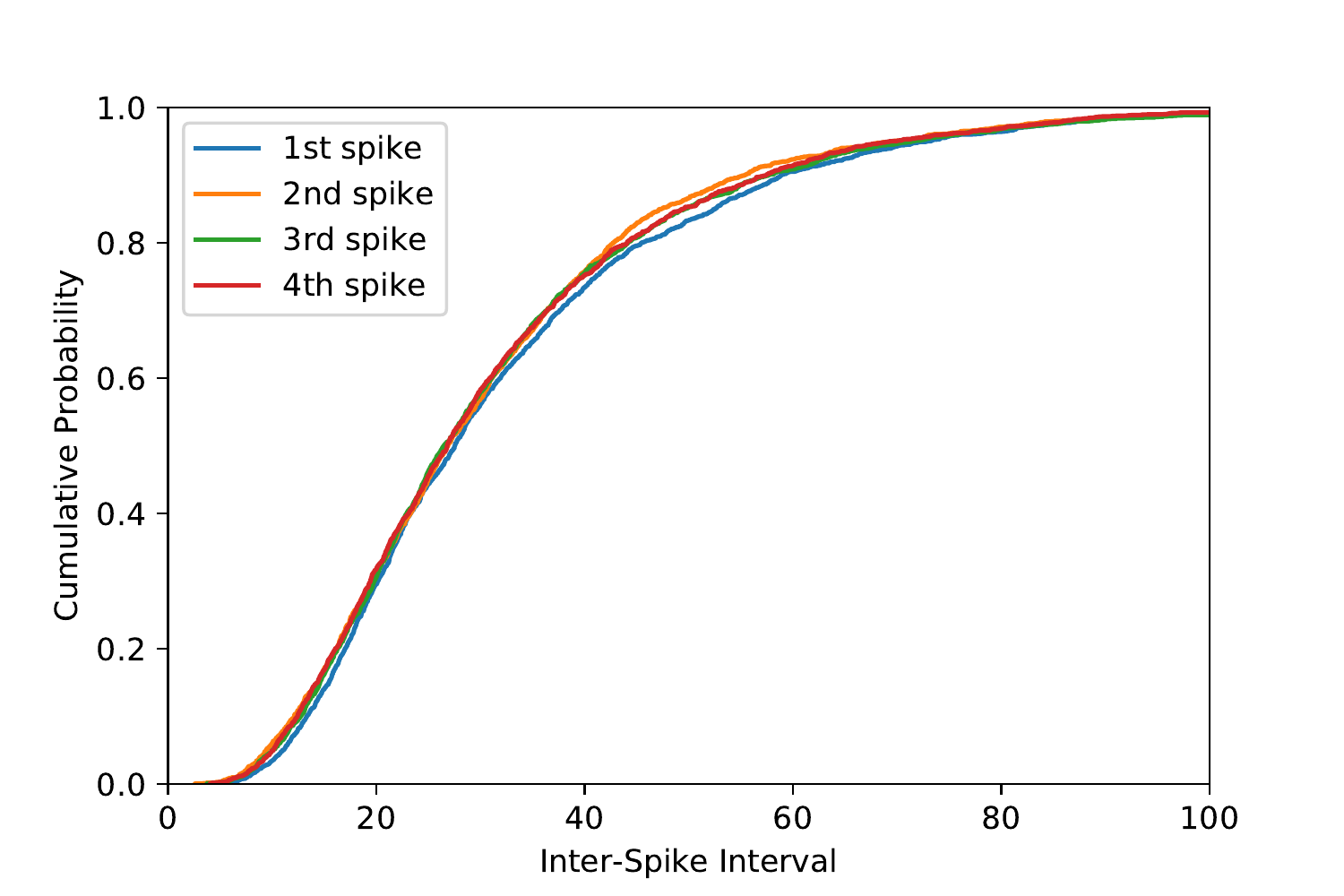}
	\end{subfigure}%
	\begin{subfigure}{0.49\textwidth}
		\centering
		\includegraphics[width=1\linewidth,trim={0cm, 0cm, 1cm, 0cm},clip]{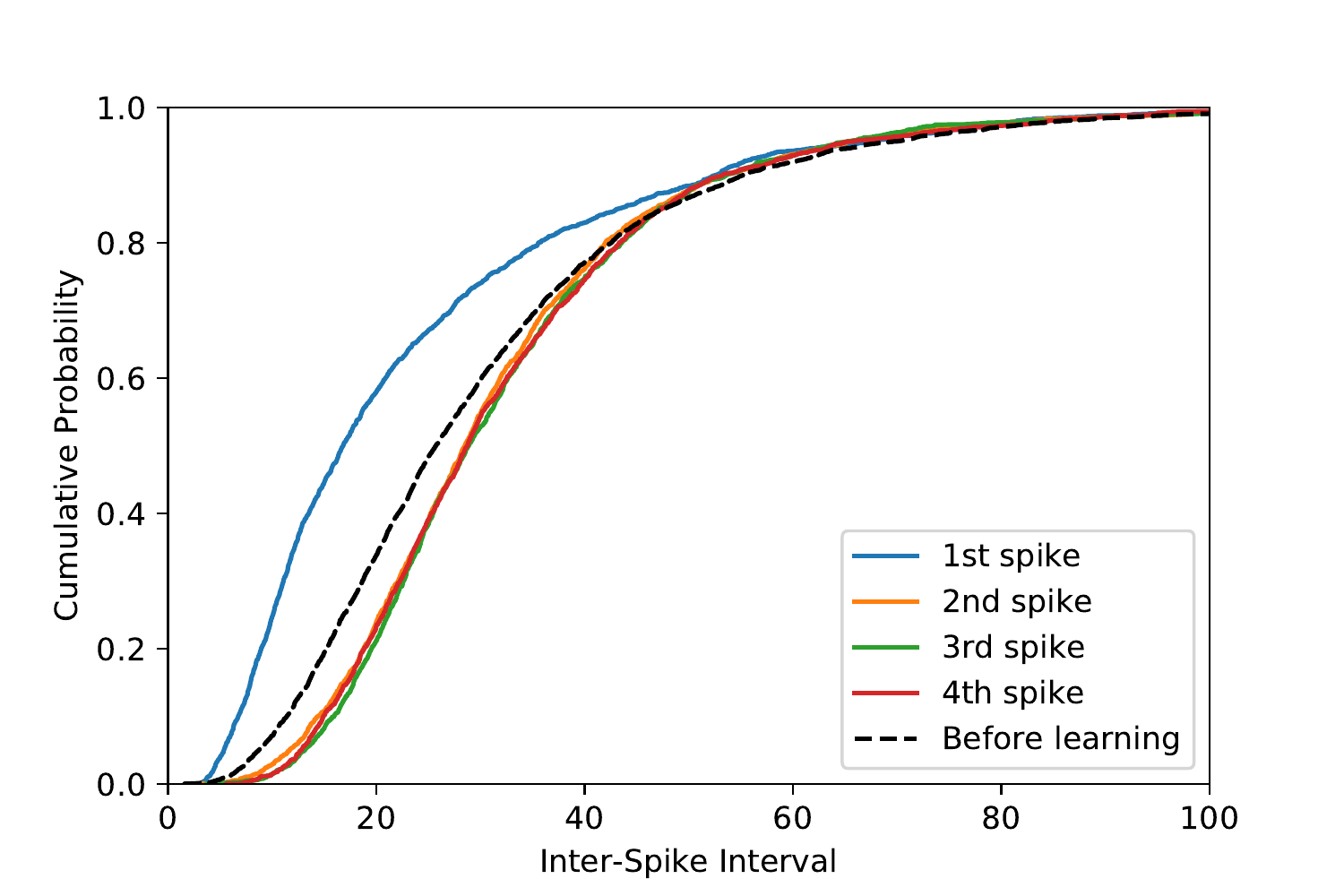}
	\end{subfigure}
	\caption{\textbf{Cumulative probability distribution of the inter-spike interval }: We plot the cumulative probability distribution of the time of the first spike and the inter-spike interval when a presynaptic spike train is presented for the first time (left) and after many repetitions (right). We simulate $2000$ neurons each receiving a presynaptic spike train lasting $600ms$ with $200$ presynaptic spikes, both inhibitory and excitatory, and whose arrival time is uniformly sampled. Every synapse evolves through STDP and being subject to both the fixed spike train with probability $0.33$ or a random pair of pre- and post-synaptic spikes with probability $0.66$. We plot the time of the first spike (blue) and the inter-spike interval for second third and fourth spikes, but subtracting the refractory period to have a pertinent comparison with the first spiking time. We can see that initially the first spike time is the same as the inter-spike interval for all the spikes, but after STDP is applied the time of the first spike reduces, implying that the blue line moves to the left with respect to the time before learning (in the black dotted line) while the other inter-spike intervals increase, thus moving the curves to the right.}\label{fig:cumulativeDistributionOfISI}
\end{figure}

After the input has been repeated many times, the distribution of postsynaptic spikes changes to $f_\infty^1$ and $f_\infty^{ISI}$ respectively. Specifically, the first spikes reduce their latency and thus move closer to $t=0$, while the inter-spike intervals increase, due to the depressive effect of postsynaptic spikes that repels or eliminates late postsynaptic spikes. Therefore,
\begin{equation}
\begin{aligned}
F_\infty^{1}(t) &= \int_{0}^{t} f_\infty^{1}(t)  dt \geq \int_{0}^{t} f_0^{1}(t) = 	F_0^{1}(t)\\
F_\infty^{ISI}(t) &= \int_{0}^{t} f_\infty^{ISI}(t)  dt \leq \int_{0}^{t} f_0^{ISI}(t) = 	F_0^{ISI}(t)
\end{aligned}
\end{equation}
where $F_\infty^{F}, F_\infty^{ISI}, F_0^{ISI}$ and $F_0^{F}$ are the cumulative probability distributions of the inter-spike intervals and first spikes respectively. This is illustrated in Fig.~\ref{fig:cumulativeDistributionOfISI} showing that indeed the first spikes move forward through STDP and the later spikes are more separated, which is consistent with the results from previous sections.

For the next section it will be convenient to look at the instantaneous firing rate, which is obtained by accumulating the times of all spikes.
\begin{equation}
s(t) = \lim\limits_{\Delta t\rightarrow 0} \sum_{k=1}^{\infty} \dfrac{\text{Pr}\left[t_k\in\left[t,t+\Delta t \right]\right]}{\Delta t}
\end{equation}
where $t_k$ is the time of the $k$th spike. Since the time of the $k$th spike is the sum of the inter-spike intervals of the first $k-1$ spikes and the first spike, and the probability of a sum is given by the convolution of the probability distributions, we can rewrite the previous function as
\begin{equation}\label{eq:instantaneousRate}
s(t) = \left(f^1 + f^1\ast f^{ISI} + f^1\ast f^{ISI} \ast f^{ISI} + ...\right)(t) 
= \left(f^1\ast \sum_{k=0}^{\infty}\left(f^{ISI}\right)^{\ast k} \right)
\end{equation}
where $\ast$ is the convolution operator, $^{\ast k}$ is the convolution power. Note that $f^1$ and $f^{ISI}$ depend on how many times the input has been repeated. We will refer to the subindex $0$ and $\infty$ to refer respectively to the cases where the presynaptic spike train is presented for the first time or when it has been presented many times.

\section{Implications for the Neural Code}\label{sec:EncodingEffects}

The postsynaptic spike trains generated by neural populations are analogous to codes that transmit information about presynaptic spikes to other neurons. As STDP is a learning mechanism that modifies the postsynaptic spike train, we expect that it should improve this encoding. Each input stimulus triggers spikes in a certain neural population, and every neuron in that population has a certain performance associated to it, the two most common ones being the energy consumption and resistance to noise \cite{rappaport1996wireless}.

\subsection{Metabolic Efficiency}\label{sec:metabolicEfficiency}

The energy consumption for a particular stimulus can be translated to our set-up as the number of postsynaptic spikes triggered when the stimulus is present. That is, we consider the integral
\begin{equation}\label{eq:SpikeCountIntegral}
	S = \int_{0}^{T} s(t) dt.
\end{equation}
When the spike train is long, we can ignore the initial distribution of spikes and use the approximation
\begin{equation}
	S\approx N_{post}\dfrac{T}{\hat{t}^{ISI}}
\end{equation} 
where $N_{post}$ is the number of postsynaptic neurons and $\hat{t}^{ISI}$ is the average inter-spike interval given by
\begin{equation}
\hat{t}^{ISI} = \int_{t_{ref}}^{\infty} t f^{ISI} dt.
\end{equation}
And by noting that
\begin{equation}
	\int_{0}^{t} f_\infty^{ISI}(t)  dt \leq \int_{0}^{t} f_0^{ISI}(t) \Rightarrow 
	\int_{0}^{T} tf_\infty^{ISI}(t)  dt \leq \int_{0}^{T} tf_0^{ISI}(t) \Rightarrow
	\dfrac{1}{\hat{t}^{ISI}_0} \leq \dfrac{1}{\hat{t}^{ISI}_\infty},
\end{equation}
therefore,
\begin{equation}
	S_\infty = N\dfrac{T}{\hat{t}^{ISI}_\infty} 
	 \leq
	 N\dfrac{T}{\hat{t}^{ISI}_0} = S_0,
\end{equation}
and thus the number of spikes decreases when the input is repeated, as illustrated in Fig.~\ref{fig:ratesAndCoding}.

\subsection{Decoding Accuracy}\label{sec:SignalToNoiseRatio}

The activity presented in a neural population must be decoded in subsequent layers of the neural hierarchy. Here we show that, under some plausible assumptions on the decoding populations, the increased concentration of spikes can make the decoding more efficient in terms of signal strength.

The first step is to consider the decoder architecture. A very generic set-up consists of a filter with a threshold detector described by the equations
\begin{equation}
\begin{aligned}
	&y(t) = \left(h\ast s\right)(t)\\
	&d = \begin{cases}
	1 \iff \max\limits_t y(t) \geq \theta
	\\
	0 \text{ otherwise},
	\end{cases}
\end{aligned}
\end{equation}
where $d$ is the decoder output, $h$ is the filter and $\theta$ is the threshold. Note that this threshold must be selected in such a way that the detection happens when the input is present -- thus low enough to be reached even in the presence of noise -- and at the same time it should not fire when the input is not present, even if there is spurious activity -- thus high enough to avoid false positives. Regardless of the choice of $\theta$, the performance of the decoder would change depending on the maximum value of $y(t)$, with higher values imply8ing better decoding performance. Our goal is then to identify regimes in which this might happen.

For simplicity, we will assume that the filter $h$ is a simple flat window given by
\begin{equation}
	h(t) = \Theta(t) - \Theta(t - L)
\end{equation}
with  $\Theta$ is the step function and $L$ is the window length. Therefore,
\begin{equation}
	 y_M=\max\limits_t y(t) = \int_{t-L}^{t}s(t) dt.
\end{equation}
The main parameter is thus the window length. When $L\sim T$, then $y_{\max} \propto S$, and it decreases when STDP is applied. Conversely, when $L$ is small $y(t)\propto s(t)$, thus the maximum is proportional to the maximum concentration of spikes, which is always at the beginning of the spike train and increases by STDP.

\begin{figure}[h]
	\begin{subfigure}{0.33\textwidth}
		\centering
		\includegraphics[width=1\linewidth,trim={0cm, 0cm, 1cm, 0cm},clip]{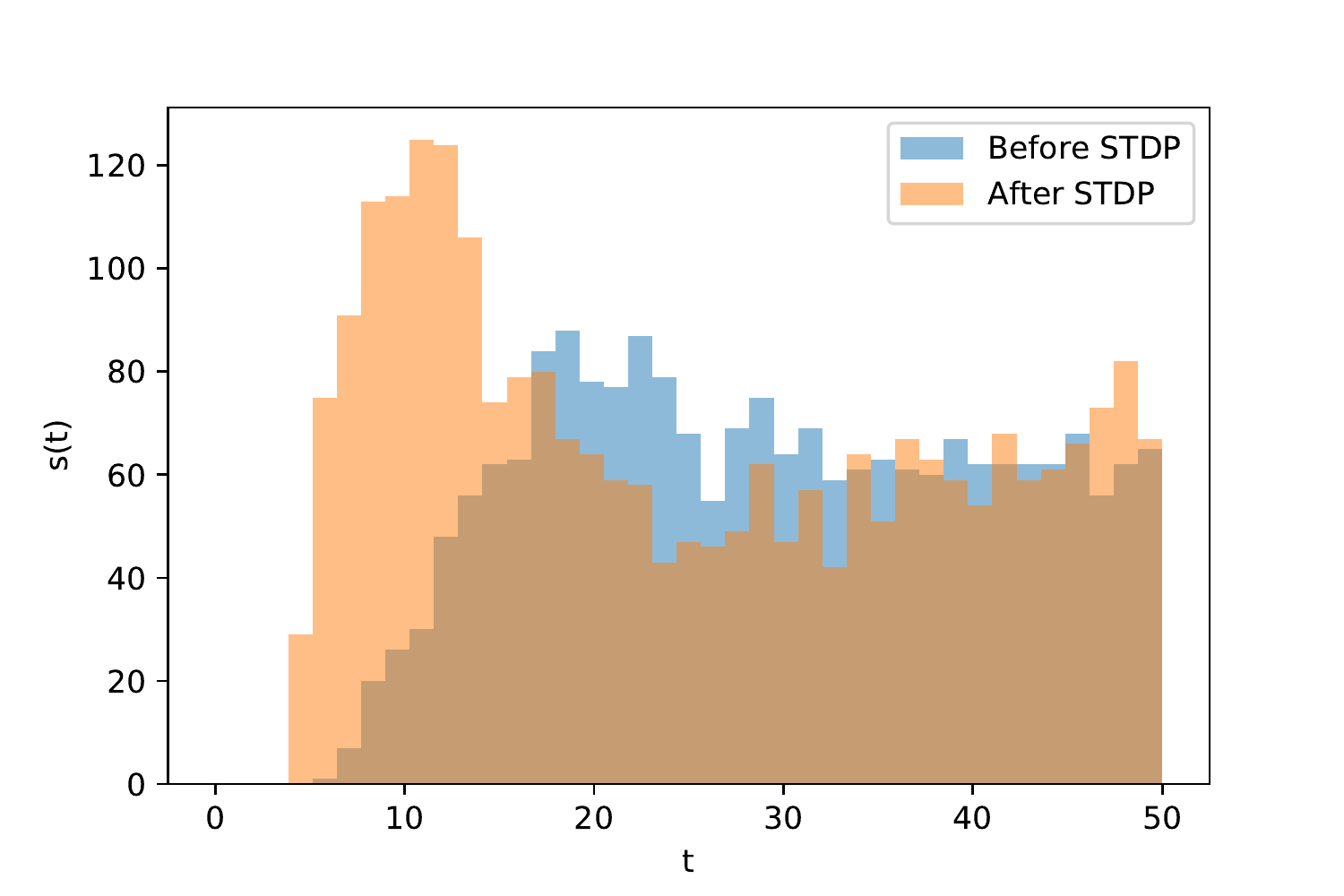}
	\end{subfigure}%
	\begin{subfigure}{0.33\textwidth}
		\centering
		\includegraphics[width=1\linewidth,trim={0cm, 0cm, 1cm, 0cm},clip]{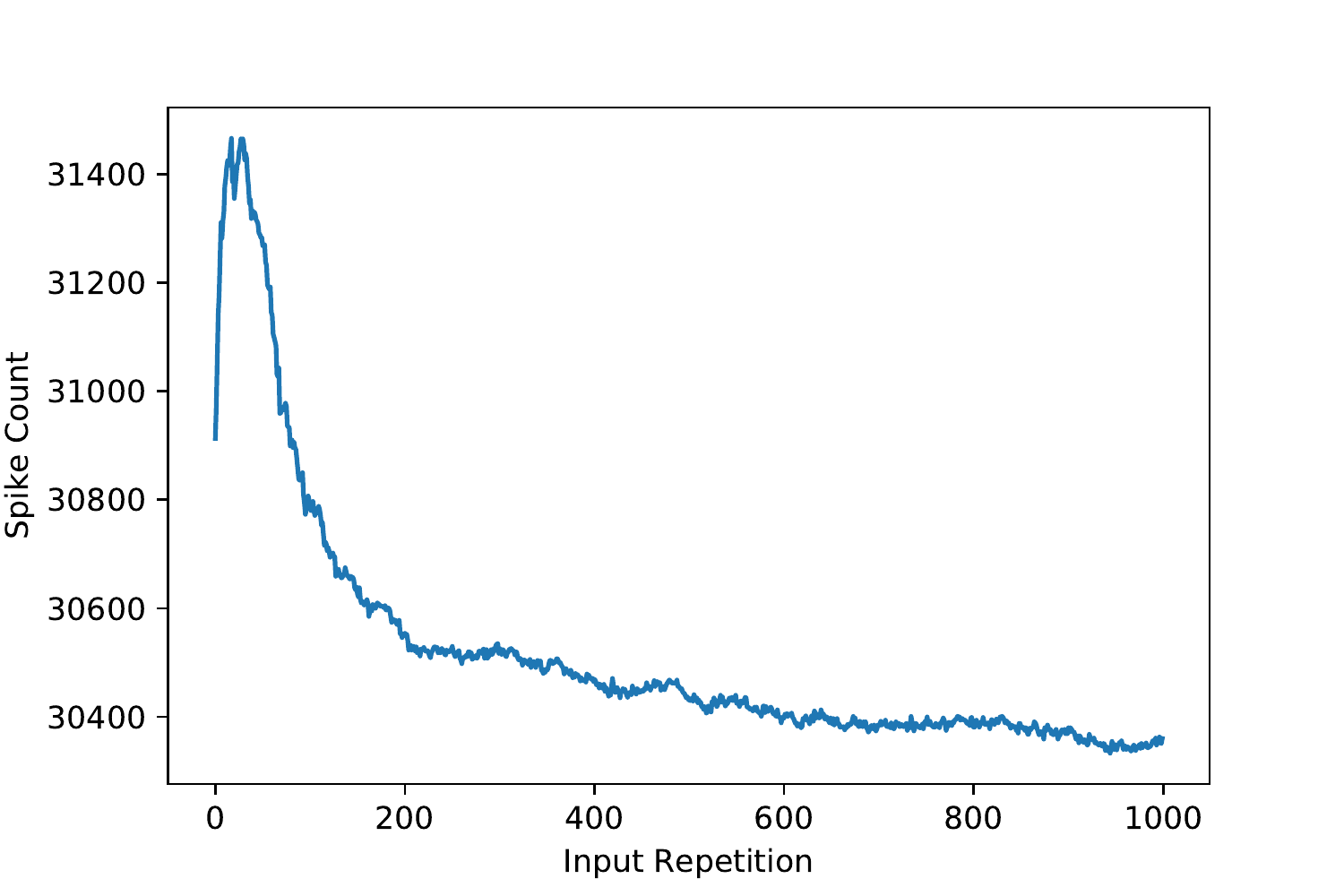}
	\end{subfigure}
	\begin{subfigure}{0.33\textwidth}
	\centering
	\includegraphics[width=1\linewidth,trim={0cm, 0cm, 1cm, 0cm},clip]{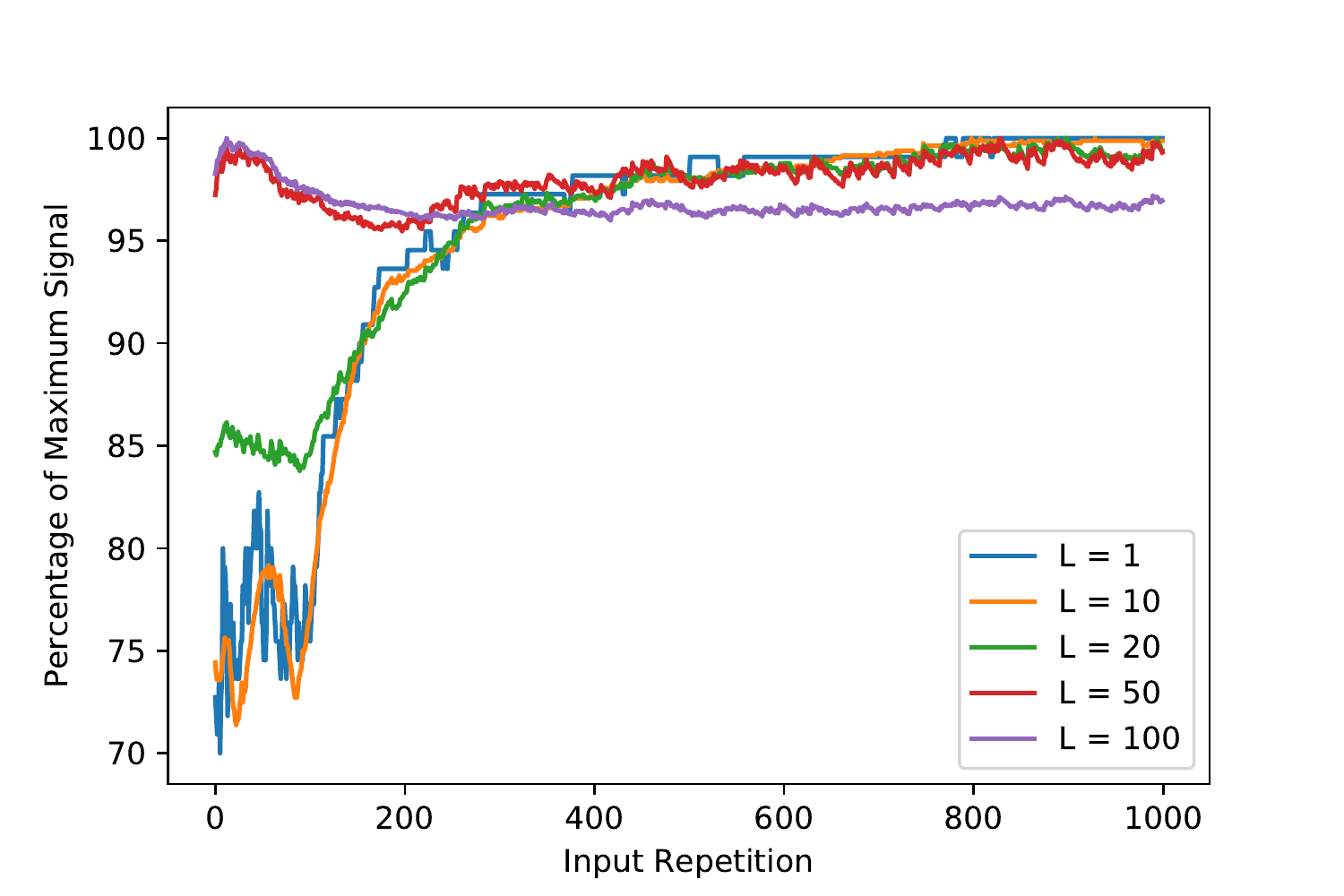}
	\end{subfigure}
	\caption{\textbf{Instantaneous firing rate and its effects on coding}: Number of spikes per bins of $4 ms$ on the first $50 ms$ of a spike train (left), evolution of the number of postsynaptic spikes per input repetition (center) and maximum signal $y_M$ per input repetition with window length of $L=\tau_m$. We simulate $1000$ neurons each receiving a presynaptic spike train lasting $600ms$ with $200$ presynaptic spikes, both inhibitory and excitatory, and whose arrival time is uniformly sampled. Every synapse is initialized by applying $500$ pairs of pre- and postsynaptic spikes with a $\Delta t$ sampled uniformly at random in the interval $[-20 ms, 20ms]$ and letting STDP modify it. Then, every synapse evolves through STDP by being subject to both the fixed spike train with probability $0.33$ or a random pair of pre- and post-synaptic spikes with probability $0.66$. As discussed in the previous sections, after the input pattern is repeated many times, the first spikes arrive early, creating an initial burst of spikes (left), but the overall number of postsynaptic spikes decreases when the input is repeated very often (center) and at the same time the initial burst of activity increases the value of $y_M$ (right) for small time windows -- $L = 1,10, 20$-- but not for long ones -- $L = 50, 100$. Notice that the reduction on the number of spikes is relatively small with respect to the original values. This should be taken as a qualitative result that depends on the parameters of the simulation, but it will increase if we had a larger time window.}\label{fig:ratesAndCoding}
\end{figure}

Thus, the final question is which value of $L$ is relevant. We consider $L=\tau_m$, which accounts for the fact that the decoder is also built from neurons which have roughly the same timescales as the encoding neurons. This implies that as long as the concentration of spikes at the beginning of the spike train is on that order of magnitude we will have an increase in $y(t)$ and thus the decoding would become better after STDP is applied, and it is indeed what we observe in Fig.~\ref{fig:ratesAndCoding}.

\section{The Emergence of Predictions}\label{sec:Predictions}
When a group of neurons encodes a stimulus we mean that those neurons fire when the stimulus is presented. However, the neurons themselves are not  aware agents and do not know anything about that stimulus; they simply receive a spike train that is strong enough to trigger their spiking. From the point of view of an encoding neurons, there is no difference between the stimulus-induced presynaptic spike train and any other input spike train that always precedes the stimulus. 

\begin{figure}
	\begin{subfigure}{0.5\textwidth}
		\centering
		\includegraphics[width=1\linewidth,trim={1.cm, 19cm, 6.2cm, 1.cm}, clip]{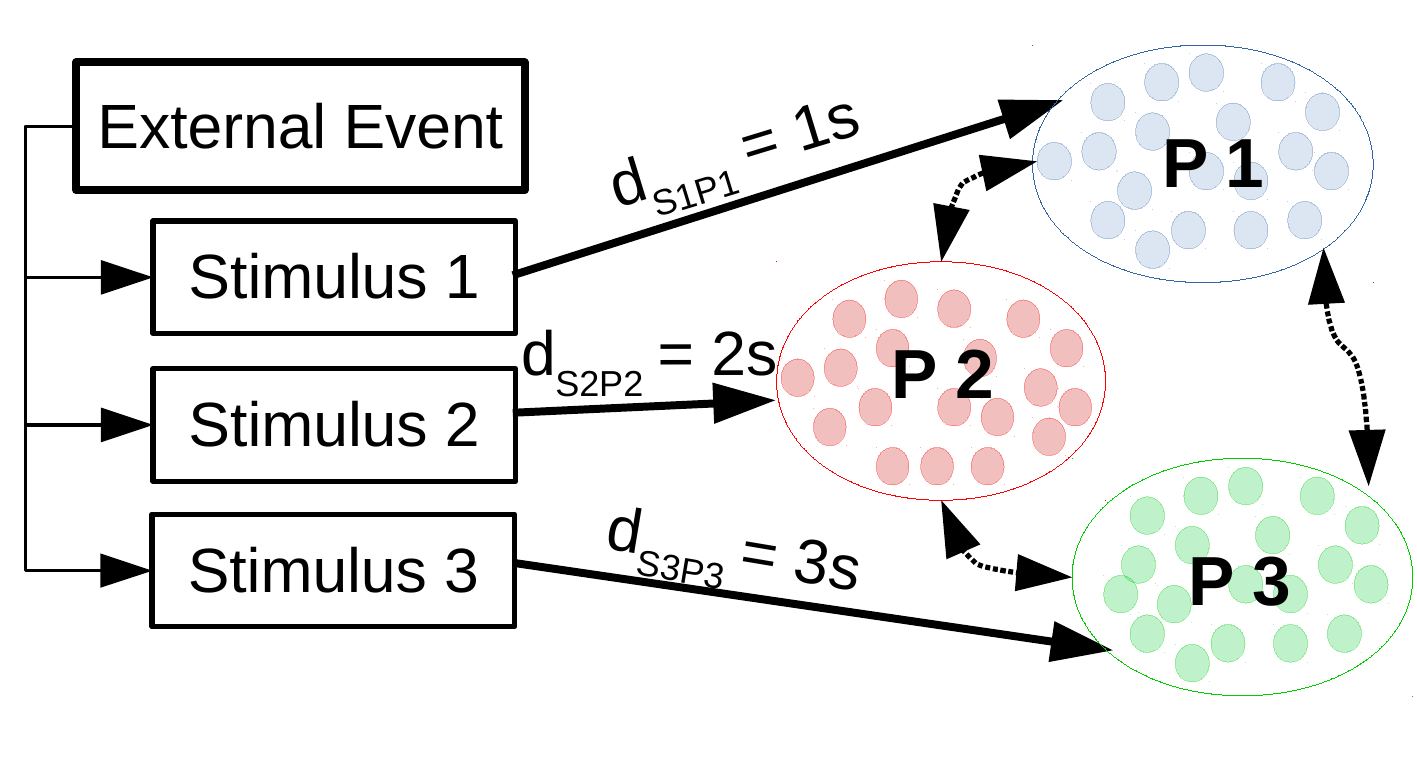} 
	\end{subfigure}
	\begin{subfigure}{0.5\textwidth}
		\centering
		\includegraphics[width=1\linewidth,trim={0cm, 0cm, 1.3cm, 0cm}, clip]{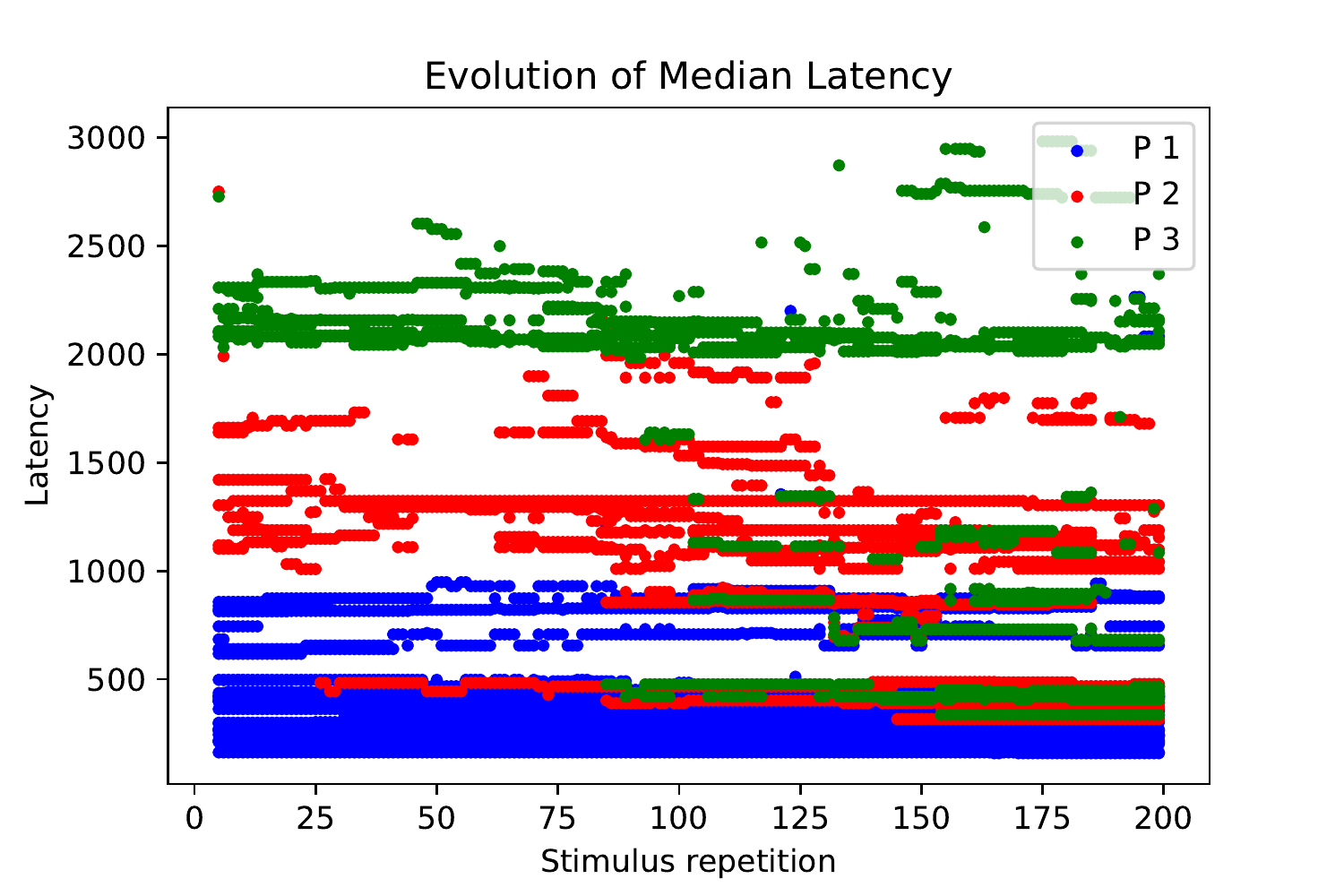}
	\end{subfigure}
	
	\caption{\textbf{Encoding Predictions}: Schema for the emergence of predictions (left) and firing latencies of neurons in encoding population (right): An external event creates three stimulus that trigger all the neurons in corresponding distinct neural populations $P1, \ P2,\ P3$ with delays randomly sampled on the intervals $d_{S1P1} \in [0ms,1000ms]$ and $d_{S2P2} \in [1000ms,2000ms]$ and $d_{S3P3} \in [2000ms,3000ms]$ respectively. The three populations, with $N=50$ neurons each one, also have synapses that between them with delays sampled from a uniform distribution between $d_{PiPj} \in [5ms,45ms]$. Originally, almost all neurons in each population fire only after receiving inputs from their respective stimuli, but after the external event is repeated very often, the inter-population connections become strong enough to trigger some spikes before the stimulus is received. Notice that even though the connections between populations are originally symmetric -- $Pi$ is connected to $Pj$ in the same way as $Pj$ is to $Pi$-- after synapses have adapted they follow the temporal order.}\label{fig:predCoding}
\end{figure}

Combining this observation with the results from previous sections showing that neurons will fire at the onset of a frequent input spike train, we can conclude that a neuron that "encodes" a stimulus can start firing before the stimulus is presented if another stimulus appears before it.
As an illustrative example, imagine listening to a melody. Different parts of the melody trigger the activity of different groups of neurons in the same area of the brain. If the melody is repeated very often, the neurons $P1$ that react to an early part of the melody will systematically fire before the neurons $P2$ that react to a later part. As the melody is repeated, neurons in $P2$ will always fire after receiving spikes from neurons in $P1$ and thus the synapses from $P1$ to $P2$ will be reinforced. Eventually, the reinforced synapses might trigger spikes in $P2$ before the late part of the melody sounds. This can be extended to more populations encoding more stimuli, and thus the whole melody is encoded through simultaneous activity of all the neurons which originally encode only separate notes. This is illustrated and simulated in Fig.~\ref{fig:predCoding}.

\section{Discussion}

In this paper we start by analyzing and expanding previous findings on latency reduction \cite{song2000competitive,guyonneau2005neurons}. Then we interpret them in communication terms: those mechanisms lead to encoding the more common inputs with less spikes while concentrating the remaining spikes in smaller time windows. This leads us to the conclusion that STDP reduces the amount of spikes used to encode frequent stimuli, in line with the idea that metabolic efficiency is one of the guiding principles of the brain \cite{hasenstaub2010metabolic, laughlin2001energy}. The same phenomena also improves decoding performance of the neural code by concentrating encoding spikes on small time windows, in line with the notion that synchronization is a learned behavior used to improve communication between neuronal assemblies \cite{singer2011dynamic,fries2005mechanism,von1994correlation}. Finally, we show that the latency reduction can explain how the nervous system learns to forecast even without any feedback. 

This study is another example of how simple, well-known plasticity rules that are present at synaptic level lead to modifications that are advantageous at the organism level. Furthermore, the fact that the same mechanism improves the neural code and creates predictions might explain how the ability of the brain to make predictions --which is one of the core problems in cognitive science-- could have emerged as a consequence of evolutionary pressures on metabolic cost and information transmission. 

Naturally, our work is also interesting for researchers in machine learning, as it shows that Hebbian learning rules, which are classically used to infer or reinforce correlations \cite{dayan2001theoretical}, can be used to find predictions by adding a temporal asymmetry in the synaptic plasticity kernel. Furthermore, the fact that the same mechanism gives rise to predictions and coding efficiency is another example of the intimate relationship between machine learning and coding \cite{mackay2003information}, thus it might be interesting for information theorists.

The results exposed here also open new questions. The effects of latency reduction in networks of neurons -- specially recurrent ones --, or the potential use of this prediction capabilities of STDP for machine learning require further study but could be useful extensions. However, the most immediate question is whether this unsupervised process is used in the nervous system. An experimental study should identify the neurons that encode two temporally correlated stimuli and follow the evolution of latencies as the stimuli are repeated, while simultaneously ensuring that this process was due to STDP alone without interference of reward systems that have been previously proposed. 

\clearpage

\appendix

\section{Postsynaptic spikes evolve independently}\label{app:uncorrelatedSpikes}

The problem with having multiple spikes per presynaptic neuron is that all of the presynaptic spikes coming from the same synapse have the same weight, and therefore when a postsynaptic spike is close to one of those presynaptic spikes, all of the presynaptic spikes that come from that synapse will undergo the same weight modifications. There are two scenarios when this would be a problem:
\begin{enumerate}
	\item A single synapse undergoes STDP from two or more different spikes: If there are two postsynaptic spikes, affected by their respective presynaptic spikes, but some of those presynaptic spikes come from the same synapse, the resulting weight change from STDP would be a combination of the effects of both postsynaptic spikes. This is undesirable as the effects could be opposite: one postsynaptic spike could induce depression while the other potentiation, and thus the evolution of one of the presynaptic spikes would not evolve as our STDP rule predicts.
	\item A new postsynaptic spike appears spontaneously from STDP: Typically, STDP applies only when there exists a postsynaptic spike. However, if some synapses are very strong due to STDP, and those synapses have spikes that are close together, they could generate a new postsynaptic spike. This would automatically generate pairs of presynaptic spikes that are affected by two postsynaptic spikes simultaneously (thus we would be in the previous case). Furthermore, the spontaneous generation of new postsynaptic spikes is itself problematic.
\end{enumerate}

Consider $M$ presynaptic neurons which fire with a rate $\lambda$, and a postsynaptic neuron that receives them with a rate $\rho = M\lambda$ during a time interval of length $T$, generating $s_{post}$ postsynaptic spikes. Furthermore, each one of those postsynaptic spikes imposes STDP that affects the presynaptic spikes that are close to it. For simplicity, we will assume that the noticeable effect on the presynaptic spikes is restricted to a time window of size $l\tau_S$ where $l$ is a small integer number. 

We start by studying case (1). If we have $s_{post}$ postsynaptic spikes, then the effects of STDP are noticeable for 
\begin{equation}
t_a = l\tau_S s_{post}
\end{equation}
milliseconds in which all presynaptic spikes should come from different synapses. Given that the arrival times of each spike are uniform of the whole interval, the expected number of presynaptic neurons that fire in that interval more than once is given by
\begin{equation}
N\sum_{k=2}^{\infty} \dfrac{(\lambda t_a)^ke^{-\lambda t_a}}{k!}= N\left(1 - e^{-\lambda t_a} - \lambda t_a e^{-\lambda t_a}\right),
\end{equation}
and by a Taylor expansion to order two,
\begin{equation}
\mathbb{E}\left[\#1\right]\approx N \left( 1- \lambda t_a + \dfrac{\lambda^2 t_a^2}{2}+ \lambda t_a - \lambda^2 t_a^2 \right) = 
N \dfrac{\lambda^2 t_a^2}{2}.
\end{equation}
To get an intuition of the magnitude of these numbers, consider, for instance, an input spike train lasting $1s$ with presynaptic spike rate of $0.5 Hz$ which generates two postsynaptic spikes and we pick the relevant time window to be twice $\tau_S$, so $l=2$ and $s_{post} = 3$. Then, the expected number of events of type (1) would be
\begin{equation}\label{eq:uncoupledBoundNeighbours}
\mathbb{E}\left[\#1\right] \approx \dfrac{M}{400}.
\end{equation}
Furthermore, not all of those events would actually be problematic; if
all of them are potentiating or depressing, then this does not change our analysis.

For case (2) we argue that in order to spontaneously generate new spikes, the synapses affected STDP must be very strong and excitatory, and a few of those strong excitatory synapses must coincide within a small time window of order $\tau_m$. 

The synapses that can be very strong are those in the $t_a$ time, meaning that we expect 
\begin{equation}
n_a = \rho t_a = \rho l\tau_S s_{post},
\end{equation}
independent synapses to be close to $w_{\max}$. Each one of those synapses can fire within the remaining $T-t_a$ time at a rate $\lambda$, so we would expect to have presynaptic a rate of STDP-affected spikes of
\begin{equation}
\lambda_{a} = n_a \lambda.
\end{equation}
Now we must compute the probability that enough of them coincide to generate a postsynaptic spike. 

We denote this number by $k$ and we will compute the number of spontaneous postsynaptic spikes that would appear for every $k$. We start by considering $k=2$ of those presynaptic spikes (although for some choices of $w_{\max}$ we have to start at $k>2$), and note that in order to have the postsynaptic spike, we must have
\begin{equation}
w_{\max} + w_{\max}e^{-\frac{\Delta t_{k=2}}{\tau_m}} + \sigma_v > v_{th} 
\end{equation}
where $\sigma_v$ is a term that accounts for the presence of other spikes that could be driving the membrane potential higher, and $\Delta t_{k=2}$ is the time interval between the two spikes. By rearranging, 
\begin{equation} 
\Delta t_{k=2} < i_{2} = \tau_m \ln\left(\vartheta - 1\right),
\end{equation}
where $\vartheta = \frac{v_{th} - \sigma_v}{w_{\max}}$. Since the spikes follow a Poisson distribution, the probability of a time interval between spikes is given by an exponential distribution, so
\begin{equation}
\text{Pr}\left[\#2|k=2\right] = \text{Pr}\left[\Delta t_{k=2} < i_{2}\right] = 1 - e^{-\lambda_a i_2},
\end{equation}
and the number of those intervals tends to $\lambda_a T$ for large $T$, so
\begin{equation}
\mathbb{E}\left[\#2\right]_{k=2} = \lambda_a T \left(1 - e^{-\lambda_a i_2}\right),
\end{equation}
For $k>2$, the estimation can be done by applying the fact that two contiguous spikes are independent, and therefore the inter-spike intervals are also independent, so we can multiply their probabilities. Furthermore, we should not have any two spikes at a distance closer than $i_2$, so
\begin{equation}
\text{Pr}\left[\#2|k=3\right] < 
\int_{i_2}^{\infty} \lambda_a e^{-\lambda_ax} \int_{i_2}^{\infty} \lambda_a e^{-\lambda_ay} \Theta\left[1 + e^{-\frac{x}{\tau_m}} + e^{-\frac{y}{\tau_m}} - \vartheta\right] dy dx,
\end{equation}
where the inequality comes because we let the interval time go to infinity, while $T$ is finite. We can ignore the $\Theta\left[1 + e^{-\frac{x}{\tau_m}} + e^{-\frac{y}{\tau_m}} - \vartheta\right]$ term and we obtain
\begin{equation}
\text{Pr}\left[\#2|k=3\right] < \left(1-e^{-\lambda_a i_2}\right)^2.
\end{equation}
And here the number of pairs of contiguous time intervals is also lower than $T\lambda_a$, which gives us
\begin{equation}
\mathbb{E}\left[\#2\right]_{k=3} \leq T\lambda_a \left(1-e^{-\lambda_a i_2}\right)^2.
\end{equation}
Naturally, the same upper bound can be computed for any $k$, so
\begin{equation}
\begin{aligned}
\mathbb{E}\left[\#2\right]_{\forall k} &= \sum_{j=2}^{\infty} \mathbb{E}\left[\#2\right]_{k=j}\leq T\lambda_a \sum_{j=2}^{\infty}\left(1-e^{-\lambda_a i_2}\right)^{j-1}\\
&= T\lambda_a e^{\lambda_a i_2} = T\lambda_a \left(\vartheta - 1\right)^{\lambda_a}.
\end{aligned}
\end{equation}
Which will be low as long as $\lambda_a$ is low. If we have, for instance, $M=50$, $l=2$, $s_{post}=3$ and $\lambda = 0.5\ Hz$, we obtain $\lambda_a = 6 \cdot 10^{-3}$. Then, if we take $\sigma_v= w_{\max}/2$, $\vartheta = 1.5$, 
\begin{equation}\label{eq:uncoupledBoundTime}
\begin{aligned}
\mathbb{E}\left[\#2\right]_{\forall k} \approx \dfrac{3T}{1000},
\end{aligned}
\end{equation}
with $T$ being in milliseconds, this means that for an input spike train lasting half a second, generating 3 postsynaptic spikes, there would be one expected spontaneous postsynaptic spike.

The estimates from Eq.~\ref{eq:uncoupledBoundNeighbours} and~\ref{eq:uncoupledBoundTime} give a relatively low number of coupled postsynaptic spikes or spontaneous spikes. We will therefore assume, from now on, that the effects described in Sec.\ref{sec:shortTimescales} are valid and happen in every postsynaptic spike on every neuron independently of the presence of other postsynaptic spikes.

\newpage

\bibliographystyle{alpha}
\bibliography{./bibliography}

\end{document}